\documentclass[prx, twocolumn, superscriptaddress, amsmath, amssymb, floatfix, longbibliography, nofootinbib, notitlepage]{revtex4-1}
\usepackage{amsbsy}
\usepackage{amsfonts}
\usepackage{amsmath}
\usepackage{amstext}
\usepackage{amsthm}
\usepackage{amssymb}
\usepackage{graphicx}
\usepackage{epsfig}
\usepackage{enumerate}
\usepackage{epstopdf}
\usepackage{bm}
\usepackage{bbm}
\usepackage{braket}
\usepackage{color}
\definecolor{prlblue}{rgb}{0.18,0.19,0.57}
\usepackage{multirow}
\usepackage[colorlinks]{hyperref}
\usepackage{cleveref}
\usepackage{ifthen}
\usepackage{tikz}
\usepackage{graphicx} 
\usepackage{amsmath, amssymb, amsfonts, mathtools, physics, float,bm,url,tabularx,bbm,slashed,xcolor}
\usepackage[letterpaper, portrait, margin=0.7in]{geometry}
\usepackage [english]{babel}
\usepackage [autostyle, english = american]{csquotes}
\usepackage{breqn}
\usepackage{titlesec}
\usepackage{enumerate}
\usepackage{epstopdf}
\usepackage{booktabs}
\MakeOuterQuote{"}
\graphicspath{ {./images/} }
\newcommand{\avg}[1]{\langle#1\rangle}

\newcommand{\sign}{\text{sgn}}

\newcommand{\fsh}[1]{\slashed{#1}}

\begin{document}
\title{Topological Phases and Phase Transitions with Dipolar Symmetry Breaking}
\author{Amogh Anakru}
\email{apa6106@psu.edu}
\author{Zhen Bi}
\email{zjb5184@psu.edu}
\affiliation{Department of Physics$,$ The Pennsylvania State University$,$ University Park$,$ Pennsylvania$,$ 16802$,$ USA}

\begin{abstract}
Systems with dipole moment conservation have been of recent interest, as they realize both novel quantum dynamics and exotic ground state phases. In this work, we study some generic properties of 1-D and 2-D dipole-conserving fermionic models at integer fillings. We find that a dipolar symmetry-breaking phase can result in a mean-field band insulator whose topological indices can strongly affect the low-energy physics of the dipolar Goldstone modes. We study the 2-D topological phase transition of the mean-field ground states in the presence of the Goldstone modes. The critical theory resembles the 2+1$d$ quantum electrodynamics coupled to massless Dirac fermions with some crucial differences and shows a novel quantum critical point featuring a nontrivial dynamical exponent. We also discuss the analogous case of 1-D dipole-conserving models and the role of topological invariants. 
\end{abstract}

\maketitle

\section{Introduction}

Generalized symmetries have recently emerged as a new paradigm in both condensed matter and high energy physics, with many profound consequences \cite{mcgreevy_generalized_symmetries}. One such example of a generalized symmetry is multipolar symmetry, i.e. the conservation of certain multipolar moments of a conserved charge density \cite{Gromov_Multipole}. Systems with multipolar symmetry have been shown to exhibit fracton-like behavior and exotic quantum dynamics\cite{fractonreview1,fractonreview2,sala2020_ergodicity_from_dipole_cons,aidelsburger_fragmentation_tilted,fractonhydro1,fractonhydro2,fractonhydro3,subdiffusion1,subdiffusion2,subdiffusion3,freezing,khemani_hermele_nandkishore_tilted,ogunnaike2023unifying}, and such symmetries have been realized experimentally in tilted optical lattices of cold atoms \cite{khemani_hermele_nandkishore_tilted,huse_bakr_tilted_lattice,scherg_aidelsburger_2021observing}. From the perspective of equilibrium physics, systems that spontaneously break their multipole symmetry have been shown to include various exotic phases of matter that defy conventional understanding in condensed matter physics. Examples include a modified Mermin-Wagner theorem for the ordering of multipolar symmetries \cite{MultipolarMerminWagner}, superfluids without a Meissner effect \cite{LakeDBHM}, and Fermi surfaces without quasiparticles \cite{anakru2023nonfermi,lakeDFHM}.


In this work, we expand our landscape of unconventional dipole-symmetry broken phases by examining fermionic models with dipolar symmetry at integer fillings. In these scenarios, breaking dipolar symmetry can result in mean-field fermion band insulators characterized by nontrivial topological indices. We investigate the relationship between mean-field band topology and the Goldstone mode arising from dipolar symmetry breaking. Previous research\cite{May-Mann_Hughes_Topo_Dipole_Cons,lakeDFHM,anakru2023nonfermi} has shown that dipolar Goldstone modes in fermionic systems mirror the spatial aspects of a $U(1)$ gauge field. Drawing on this analogy, we anticipate that the topological nature of the band structure will introduce intriguing topological terms into the Goldstone action. A key objective of this paper is to understand how these topological terms influence the low-energy behaviors of the Goldstone modes.  Furthermore, as microscopic parameters within the model undergo changes, the mean-field band structure may experience topological transitions. We aim to explore how the universal characteristics of such transitions are affected by the strong interaction with the dipolar Goldstone modes. It is worth mentioning that the intersection of multipole symmetries and ground-state topology has previously been investigated, revealing that multipole symmetry can lead to novel forms of gapped insulators, filling constraints, and symmetry-protected topological orders\cite{May-Mann_Hughes_Topo_Dipole_Cons,burnell2023filling,han_lake_dipolar_chains,lake_Dipolar_SPT,ho_tat_lam_dipolar_SPT_classification,ho_tat_topo_dipole_insulator_2024}. Our current study focuses on a distinct set of systems and phase transitions, characterized by the spontaneous breaking of dipolar symmetry and the prominent role of dipolar Goldstone modes. 

To set the stage for the rest of our work, we discuss in Sec. \ref{Sec:Pure1D} a simple 1-D setting where topological considerations can shed light on properties of dipolar-ordered states and the transitions between such phases. In Sec. \ref{Sec:Toy_Model}, we construct specific dipole-conserving models whose mean-field states include insulators with various Chern numbers. We then shift our focus to effective field theories for the remainder of the work. Sec. \ref{Sec:Goldstone_EFT} derives the effective field theory for dipolar Goldstone modes within the ordered phases found in the previous section, featuring the effect of nontrivial topological terms generated by the topological band structures. Finally in Sec. \ref{Sec:QHTransition}, we discuss the transition between mean-field states with different Chern numbers, which generalizes the quantum Hall transition as studied in \cite{mpa_fisher_mott_transition_anyon_gas} but with stark differences. We explore the critical properties of this transition within a $1/N$-expansion by self-consistently computing an effective action for the Goldstone and fermion propagators. The critical theory will be strongly renormalized because of the coupling between the fermions and Goldstone modes and presents an intriguing interaction dominated scaling\cite{schlief_lunts_lee_exact_critical_afm,lunts_schlief_lee_emergence_control_afm,sung_sik_NFL_review}.

\section{1-D Topological Phases with Dipolar Symmetry Breaking}
\label{Sec:Pure1D}

Before discussing the relationship between ground-state topology and dipolar symmetry breaking in 2-D, we first focus on the conceptually simpler case of 1-D dipole-conserving models. First, we present a 1-D toy lattice model of fermions at integer fillings. The Hamiltonian is defined on a 1-D chain with two orbitals $c_i,d_i$ per unit cell:
\begin{equation}\label{eq:App_Micro_Model_SSH}
    \hat{H}= \sum_i \biggr(h(c_i^\dag d_i+d_i^\dag c_i) - t_{d}(d_i^\dag c_{i+1})\mathcal{A}_{ij}(c_{j+1}^\dag d_{j})\biggr)
\end{equation}
where $\mathcal{A}_{ij} = \mathcal{A}_{ji}^*$.  This model has the usual $U(1)$-charge symmetry and (on an infinite or open chain) a $U(1)$-dipole symmetry generated by 
$\hat{Q}_{dip} = \sum_j\,aj(c_j^\dag c_j + d_j^\dag d_j - 1)$ where $a$ is the lattice spacing. Additionally, the model has a particle-hole symmetry $\mathcal{P}^{-1}\hat{H}\mathcal{P}$ where $\mathcal{P}$ maps $c_i\leftrightarrow c_i^\dag$ and $d_i\leftrightarrow -d_i^\dag$ and anticommutes with the dipole operator $\hat{Q}$. We consider the system at a filling of one fermion per unit cell. 

We assume that at large $t_d$ the system will prefer to form a quasi-long range order of the dipolar symmetry. We take a mean-field ansatz $\avg{\mathcal{A}_{ij}(c_{j+1}^\dag d_{j})}=\Delta$, and Goldstone fluctuations are parametrized by writing $\Delta_i =  \abs{\Delta} e^{ia\varphi_i}$. Under the $P$ symmetry, the Goldstone field transforms as $\varphi\rightarrow -\varphi$. Accordingly, the mean-field Hamiltonian with Goldstone fluctuations takes the following form,
\begin{equation}
    \label{eq:The_SSH_Model_Rotor}
    \hat{H} = \sum_i \biggr(hc_i^\dag d_i + \Delta e^{-ia\varphi} d_i^\dag c_{i+1} + h.c.\biggr) + \hat{H}_\varphi,
\end{equation}
where $\hat{H}_\varphi$ governs fluctuations of the dipolar Goldstone field. Interestingly, the field $\varphi$ couples to the lattice fermions as a spatial pseudo-gauge field, and is also compactified according to $\varphi\cong \varphi+2\pi/a$. The mean-field Hamiltonian resembles the SSH model \cite{SSH_OG}. The SSH model is described by a topological invariant $\theta = \sum_{n\in \mathrm{occ.}}\int_0^{2\pi/a} dk_x\mathcal{A}_n(k_x)$ with the Berry phase $\theta=\pi$ if $\abs{\Delta}>h$ and $\theta=0$ when $\abs{\Delta}<h$. Given the filling, the fermions are in a band-insulator. We can integrate out the fermions to get an effective field theory for the gapless Goldstone mode $\varphi$. Due to the pseudo-gauge field analog, such an effective field theory, to the quadratic order, should be given by: 

\begin{equation}\label{eq:Effective_Goldstone_SSH}
   \mathcal{L}_{eff}[\varphi^a] =\frac12\kappa (\partial_t\varphi)^2 -\frac12K(\partial_x\varphi)^2 +\frac{\theta}{2\pi}\partial_t\varphi.
\end{equation}

The coefficients $\kappa$ and $K$ describe the dipole compressibility and gradient energy, respectively, which are in general not universal and will get contributions from the fermion band energy. However, the linear term proportional to $\theta$ is a topological term which is in fact universal. The $\theta$ is precisely the Berry phase of the occupied fermion band in Eq. ~\eqref{eq:The_SSH_Model_Rotor}. This term is analogous to the term $\frac{1}{2\pi}\theta E_x$ that shows up in the effective electrodynamics of any 1-D band insulator. We could identify $\partial_t\varphi$ as the "electric field" associated to the pseudo-gauge field $\varphi$. The $\mathcal{P}$ symmetry in the system as defined previously quantizes the Berry phase to be $\theta\in\{0,\pi\}$. To see this, note that the symmetry $\mathcal{P}$ flips $\varphi\to-\varphi$ so that the total Hamiltonian Eq. ~\eqref{eq:The_SSH_Model_Rotor} remains $\mathcal{P}$-symmetric. As a result, in the effective field theory, $\mathcal{P}$ flips $\theta\to-\theta$. Due to the periodicity of the $\theta$ angle, only $\theta=0,\pi$ are compatible with the $\mathcal{P}$-symmetry. Therefore, any 1-D dipole- and $\mathcal{P}$-symmetric lattice Hamiltonian exhibiting a phase with dipolar quasi-long range order should be described at long wavelengths by Eq. \ref{eq:Effective_Goldstone_SSH} with $\theta\in\{0,\pi\}$.

We now wish to study properties of this effective field theory in Eq. ~\eqref{eq:Effective_Goldstone_SSH}. Immediately, we notice that the term $\theta\partial_t\varphi/2\pi$ is a total derivative, and thus there is no effect on the dynamics of the Goldstone mode from this term alone. This is in direct contrast to the 2-D case discussed later in Sec. \ref{Sec:Goldstone_EFT} where the topological invariant of the mean-field ground state will quite drastically change the dynamics of the Goldstone modes. Nevertheless, in the following sections, we will discuss some physical consequences of this topological term, with emphasis on the various phase transitions out of the ordering phase and between phases with different fermion band structure. 

\subsection{Robust edge modes}
\label{Sec:Edge_Modes_In_1D}
A key feature of the non-interacting SSH model on a open chain is the presence of edge modes that become degenerate in the thermodynamic limit. These edge modes are expected to persist as degrees of freedom in the effective Hamiltonian \ref{eq:The_SSH_Model_Rotor}, but these edge modes may now couple to the bulk gapless mode $\varphi$. Thus, it is not obvious that the edge degeneracy is stable on a finite chain. While any splitting of the edge modes certainly decays as some function of the system size $L$, this could be comparable to the $\sim1/L$ decay of the finite size gap for the bulk modes, which would make it impossible to distinguish the edge and the bulk modes. 

Remarkably, it turns out that the presence of particle-hole and dipole symmetries $\mathcal{P}$ and $\hat{Q}_{dip}$ prevents any splitting for the edge modes in a model like Eq. \ref{eq:The_SSH_Model_Rotor} (and generally any model described by Eq. \ref{eq:Effective_Goldstone_SSH} with $\theta=\pi$). To see this, suppose that a ground state $\ket{G}$ of the Hamiltonian $\hat{H}$ derived from Eq. \ref{eq:Effective_Goldstone_SSH} has $\hat{Q}_{dip}\ket{G} = q_{dip}\ket{G}$. Because $\mathcal{P}$ preserves $\hat{H}$, $\mathcal{P}\ket{G}$ is an eigenstate of $\hat{H}$ with the same energy. But because $\mathcal{P}^{-1}\hat{Q}_{dip}\mathcal{P} = -\hat{Q}_{dip}$, the state $\mathcal{P}\ket{G_1}$ has a dipole charge of $-q_{dip}$. Supposing $q_{dip} \neq 0$, $\mathcal{P}\ket{G_1}$ is thus orthogonal to $\ket{G_1}$, and thus the ground state manifold is at least doubly degenerate. If $\theta=\pi$, i.e. the mean-field sector is described by a fermion chain with topological edge modes, it is expected that the many-body ground state indeed has a finite dipole charge\footnote{Note that since $\hat{Q}_{dip}$ is an exact symmetry of the Hamiltonian, we are always guaranteed a ground state that is also an eigenstate of $\hat{Q}_{dip}$. This may not be the unique ground state, and per the above discussion, it is certainly not unique if the $\hat{Q}_{dip}$ eigenvalue $\neq 0$. This proof also holds for excited states as well; i.e. $\hat{H}$-eigenstates with nonzero $\hat{Q}_{dip}$ come in pairs related by $\mathcal{P}$.}. Specifically, on a finite, open chain, either one of the two fermionic edge modes can be occupied, and either one of these possibilities corresponds to a many-body ground state with $q_{dip} \neq 0$. Thus, no operator that preserves particle-hole and dipole symmetry can lift the degeneracy of these two ground states.

\subsection{Disordering transitions}

\begin{figure}
    \centering
    \includegraphics[width=\linewidth]{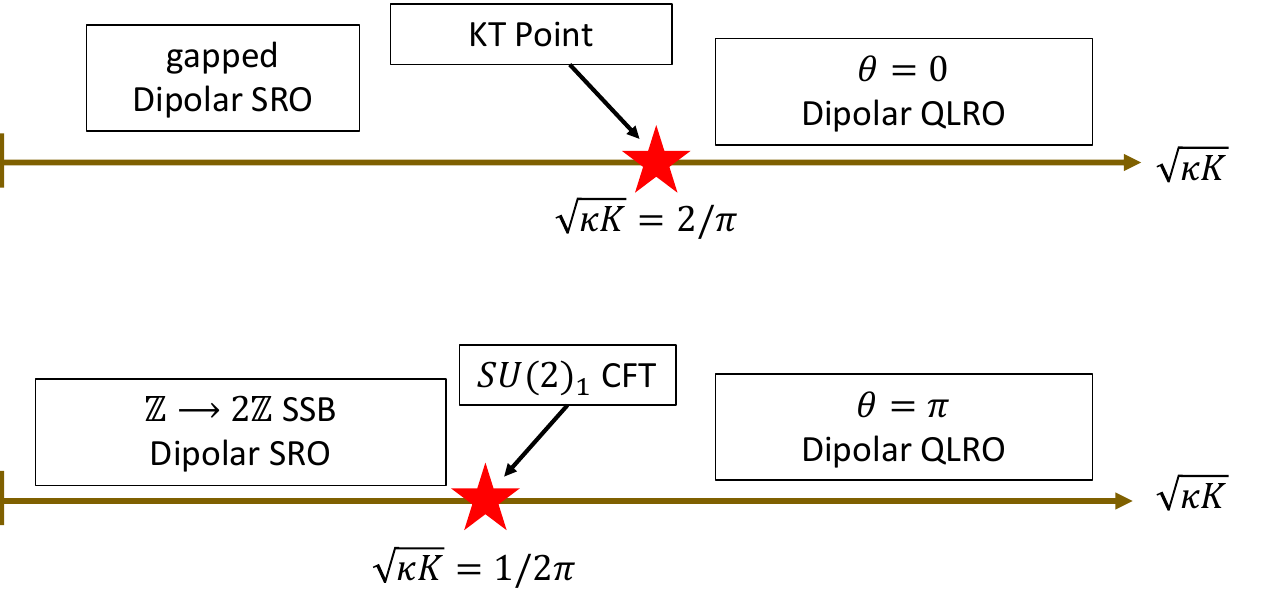}
    \caption{Phase diagrams of disordering transitions with and without the topological response.}
    \label{fig:KT_transitions}
\end{figure}

In this section, we discuss the influence of the topological term on the phase transition that destroys the quasi-long range order of the dipolar symmetry. To that end, we need to consider the proliferation of topological defects of the $\varphi$ field. The term $\frac{\theta}{2\pi}\partial_t\varphi$ notably does not influence the dispersion relation of the $\varphi$ mode, but it is felt by vortex configurations of $\varphi$. In particular, per an observation of Haldane \cite{Haldane_1D_LSM_Field_Theory_Thing_Vortex_Momentum}\footnote{See also App. A of \cite{Lake_Tilted_Chain}.}, single vortices with $\oint d\vec{s}\cdot \nabla_{x,t}\varphi(x,t)=\pm 2\pi/a$ are endowed with a momentum $\pi/a$\footnote{The field-theoretic momentum, i.e. the Noether charge associated to translation symmetry $\varphi_x(x)\to\varphi_x(x+\epsilon)$, contains a most relevant term $P(t) \approx \int dx\,\frac{\theta}{2\pi}\partial_x\varphi(x,t) = \frac{\theta}{2\pi}(\varphi_x(\infty,t)-\varphi_x(-\infty,t)$. A vortex of strength $m$ located at $(x_v,t_v)$ will thus carry $P(t>t_v)-P(t<t_v) = m\theta/a$.} and thus a $2\pi$-vortex creation operator is forbidden as a perturbation to the Gaussian theory. The next most relevant operator allowed by symmetry is a $4\pi$ vortex creation operator, which may be added to the theory because a momentum $2\pi/a \cong 0$ on a lattice. When this double-vortex operator becomes relevant, the cosine potential has two inequivalent minima corresponding to a spontaneous breaking of translation symmetry, and the critical point separating this phase from the dipolar quasi-long range ordered phase is described by an $SU(2)_1$ CFT \cite{affleck1989}. We contrast this to the usual case of the XY model at $\theta=0$ where the short-range correlated phase has a nondegenerate ground state. Accordingly, the two values $0,\pi$ of $\theta$ in Eq. \ref{eq:Effective_Goldstone_SSH} describe genuinely different gapless phases with different possibilities for their transitions into gapped phases. These two possibilities for disordering transitions are depicted in Fig. \ref{fig:KT_transitions}, as a function of the parameters $\kappa, K$.

The above discussion are on the possibilities of continuous phase transitions. There is also the possibility of a first-order transition out of the $\theta=\pi$ phase into a trivial gapped phase with short-range dipole correlations. In fact, when a standard mean-field computation is applied to Eq. ~\eqref{eq:App_Micro_Model_SSH} (for $\mathcal{A}_{ij}=\delta_{i,j+1}+\delta_{i,j-1}$), the condensate $\abs{\Delta}$ discontinuously jumps from 0 to a value greater than $h$ when $t_d/h>r_c\approx 1.68$. Thus, the only phases predicted by the mean-field theory are a gapped symmetry-unbroken phase and a gapless $\theta=\pi$ phase seperated by a first-order transition. Given that this comes from a mean-field computation in $(1+1)$-dimensions, it is not clear that such a first-order transition is actually present. Furthermore, the absence of a $\theta=0$ quasi-long range ordered phase within mean-field theory is a particularity of the specific model, and adding more symmetry-allowed terms may lead to a richer phase diagram hosting both $\theta=0$ and $\theta=\pi$ phases. Our preliminary numerical calculations with a more comprehensive model indeed show a diverse phase diagram, and we will leave this subject to a future work. 

\subsection{Transitions between ordered phases}
\label{Sec:1D_Ordered_Transitions}
Having discussed phase transitions into disordered phases, we now seek a transition between the trivial and topological phases where $\theta$ changes from $0$ to $\pi$. In order to do this, we will first write down a continuum theory which flows to Eq. \ref{eq:Effective_Goldstone_SSH} for $\theta=0,\pi$ depending on a tuning parameter, and then study this theory at the critical point. Besides the gapless Goldstone modes, the lowest-energy excitations of Eq. \ref{eq:The_SSH_Model_Rotor} are fermionic excitations near $k=\pm \pi/a$; at long wavelengths, these excitations are described by a massive Dirac fermion in $(1+1)$-dimensions. Accordingly, we can write the following continuum Lagrangian that describes both the fermion and Goldstone degrees of freedom:
\begin{equation}\label{eq:DiracAction}
\begin{split}
\mathcal{L}&=\bar{\psi}\gamma^\mu(i\partial_\mu-\varphi_\mu) \psi+\bar{\psi}me^{i\theta\gamma^5}\psi + \mathcal{L}_{\varphi}\\
\mathcal{L}_{\varphi} &= \frac12\tilde{\kappa} (\partial_t\varphi)^2 -\frac12\tilde{K}(\partial_x\varphi)^2.
\end{split}
\end{equation}
where $\psi=(\psi_{c},\psi_d)$, $\{\gamma^0,\gamma^1, \gamma^5\}=\{\sigma_y,i\sigma_x,\sigma_z\}$, $\partial_\mu=(\partial_t,\partial_x)$, $\varphi_\mu = (0,-\varphi)$, and we have assumed an appropriate rescaling of the spatial coordinate. The angle $\theta\in\{0,\pi\}$ is the Berry phase of the filled fermion band. Because the massive Dirac fermion is gapped, we may integrate it out by computing the fermion determinant and expressing it in terms of diagrams with external $\varphi$-legs, in the usual manner. The effective theory in Eq. \eqref{eq:Effective_Goldstone_SSH} can be recovered by evaluating diagrams with $1$ or $2$ $\varphi$-legs. We regularize the theory such that $m<0$ side has $\theta=\pi$ and $m>0$ has $\theta=0$.


We offer another useful perspective on this model via bosonization of Eq. \ref{eq:DiracAction}:
\begin{equation}\label{eq:DiracActionBosonized}
\begin{split}
\mathcal{S} = \frac{1}{2}\int dt dx \biggr(&(\partial_t\phi)^2-(\partial_x\phi)^2+2m\cos(2\sqrt{\pi}\phi-\theta) \\
&-\frac{2}{\sqrt{\pi}}\varphi_x\partial_t\phi + \frac{1}{g^2}((\partial_t\varphi_x)^2 - v_B^2(\partial_x\varphi)^2)\biggr).
\end{split}
\end{equation}
Here, the field $\phi$ is related to the microscopic fermions by $\epsilon^{\mu\nu}\partial_\nu\phi = \sqrt{\pi}\bar{\psi}\gamma^\mu\psi$ and we have reparametrized the constants $\kappa,K$ into $g, v_B$. Shifting $\phi\to\phi+\theta/2\sqrt{\pi}$ and integrating out the now-gapped $\phi$-field, we recover the $\theta$-term in Eq. \ref{eq:Effective_Goldstone_SSH} after a suitable field rescaling\footnote{This rescaling is chosen in order to match the compactification radii of Eqs. \ref{eq:Effective_Goldstone_SSH} and \ref{eq:DiracActionBosonized}; In Eq. \ref{eq:DiracActionBosonized}, we have $\phi\cong \phi+2\sqrt{\pi}$ and thus $\varphi_x\cong \varphi_x+\pi/a$.}.

\begin{figure}
    \centering
    \includegraphics[width=1\linewidth]{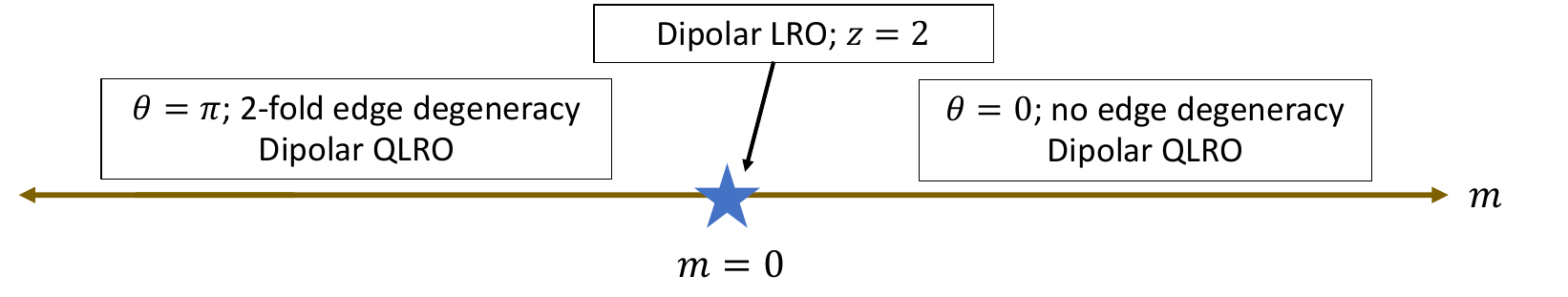}
    \caption{Phase transitions between trivial and topological band structures.}
    \label{fig:Dirac_Phase_diagram}
\end{figure}

Now, we focus on the topological transition by tuning the mass term of the Dirac fermions. The resulting quantum critical point we wish to study is the theory of a massless Dirac fermion coupled to a dipolar Goldstone mode. Such a theory has already been studied in \cite{anakru2023nonfermi} and exhibits many interesting features—importantly, the model is exactly solvable. To summarize the key results, this critical point is described by a Lifshitz theory (i.e. the dynamical exponent $z=2$) and exhibits long-range order for the dipolar order parameter $e^{i\varphi}$. Furthermore, it has a dual description in terms of the 1-D dipole-conserving superfluid as studied in \cite{Lake_Tilted_Chain,zechmann2022fractonic}. The action of this dual bosonic theory is precisely the theory Eq. \eqref{eq:DiracActionBosonized} with $m=0$. With reference to this bosonized form, there is always an instability to the cosine term $2m\cos(2\sqrt{\pi}\phi-\theta)$ no matter what the Luttinger parameters of the boson fields are. This term is dual to the Dirac mass and is the relevant parameter that tunes the phase transition. These conclusions are summarized in Fig. \ref{fig:Dirac_Phase_diagram}.


\section{2-D Topological Phases with Dipolar Symmetry Breaking}
\label{Sec:Toy_Model}
\subsection{A toy model with mean-field Chern insulator}
In this section, we study a family of dipole-symmetric models of spinful fermions on the 2-D square lattice whose mean-field phase diagram exhibits a variety of dipole-condensed phases. In the mean-field treatment, we find phases with nontrivial topological band structures and phase transitions out of the topological phases. This serves as motivation for us to study the universal features of topological phase transitions in the presence of dipolar symmetry breaking, which is the focus of the rest of the paper. 

The simple toy model is written as 
\begin{equation} \label{eq:TopologicalToyModel}
\begin{split}
\hat{H} = &-\mu\hat{N} +\sum_i\Psi^\dag_i M\Psi_i +\sum_{ij,a}(\hat{D}^a_i)^\dag\mathcal{A}^a_{ij}\hat{D}^a_j
\end{split}
\end{equation}
where $\Psi^\dag = (c^\dag_{\uparrow},c^\dag_{\downarrow})$, $(\hat{D}^a_i)^\dag  = \Psi_i^\dag T_a\Psi_{i+a}$, $M=\vec{M}\cdot\vec{\sigma}$, and $T_a = \vec{t}_a\cdot\vec{\sigma}$; $a$ ranges over the $2$ primitive unit vectors. Here, we focus on the case where
\begin{equation}\label{eq:Choice_of_Matrices}
   T_x=(iw\sigma_x+\alpha \sigma_z),\,\,\,T_y = (iw\sigma_y+\alpha\sigma_z),\,\,\,M=h\sigma_z.
\end{equation}
The Hermitian matrix $\mathcal{A}_{ij}^a$ describes the correlated hopping of dipole moments on the lattice; the operators $\hat{D}^a_j$ are dipole-creation operators that carry an integer charge under the dipole symmetry generator $\hat{Q}_{dip}^a=\sum_i(\hat{a}\cdot\vec{r}_i)\hat{n}_i$. 
For simplicity of the discussion, we have only assumed charge and dipole conservation as internal symmetries. Time-reversal and spin rotational symmetries are explicitly broken in the Hamiltonian.  

The dipolar hopping terms will prefer dipole condensation. We may assume a mean-field decoupling where $\avg{\mathcal{A}^a_{ij}\hat{D}^a_j}=\Delta^a$; if $\Delta_i^a\neq 0$, then the dipole symmetry is spontaneously broken. Accordingly, the mean-field Hamiltonian assumes a standard tight-binding form
\begin{equation} \label{eq:QuadraticMeanField}
\begin{split}
\hat{H}_{MF} = &-\mu\hat{N} +\sum_i\Psi^\dag_i M\Psi_i +\sum_{i,a}(\Delta^a(\hat{D}^a_i)^\dag+h.c.)\\
&-\sum_{ij,a}\Delta^a_i(\mathcal{A}^a_{ij})^{-1}\Delta^a_j\end{split}
\end{equation}
and we assume the system is at filling of one fermion per unit cell. For a given choice of matrices $T_a$, $M$, the mean-field Hamiltonian Eq. \ref{eq:QuadraticMeanField} generally describes a band insulator whose topology can be tuned by $T_a, M$, or the amplitude of the order parameter $\abs{\Delta^a}$. With our specific choice of matrices in \ref{eq:Choice_of_Matrices}, Eq. \ref{eq:QuadraticMeanField} is a slightly modified version of the model introduced by Qi, Wu, and Zhang \cite{qi_wu_zhang_2006} for a quantum anomalous hall state. The model is known to undergo a topological phase transition involving a change in the Chern number as $w$ and $\alpha$ are varied. We can interpret $w$ as the strength of the spin-orbit coupling and $\alpha$ as the spin-dependence of the effective mass. The parameter $h$ is akin to a magnetization for the spins.

\begin{figure}
\centering
\includegraphics[width=0.45\textwidth]{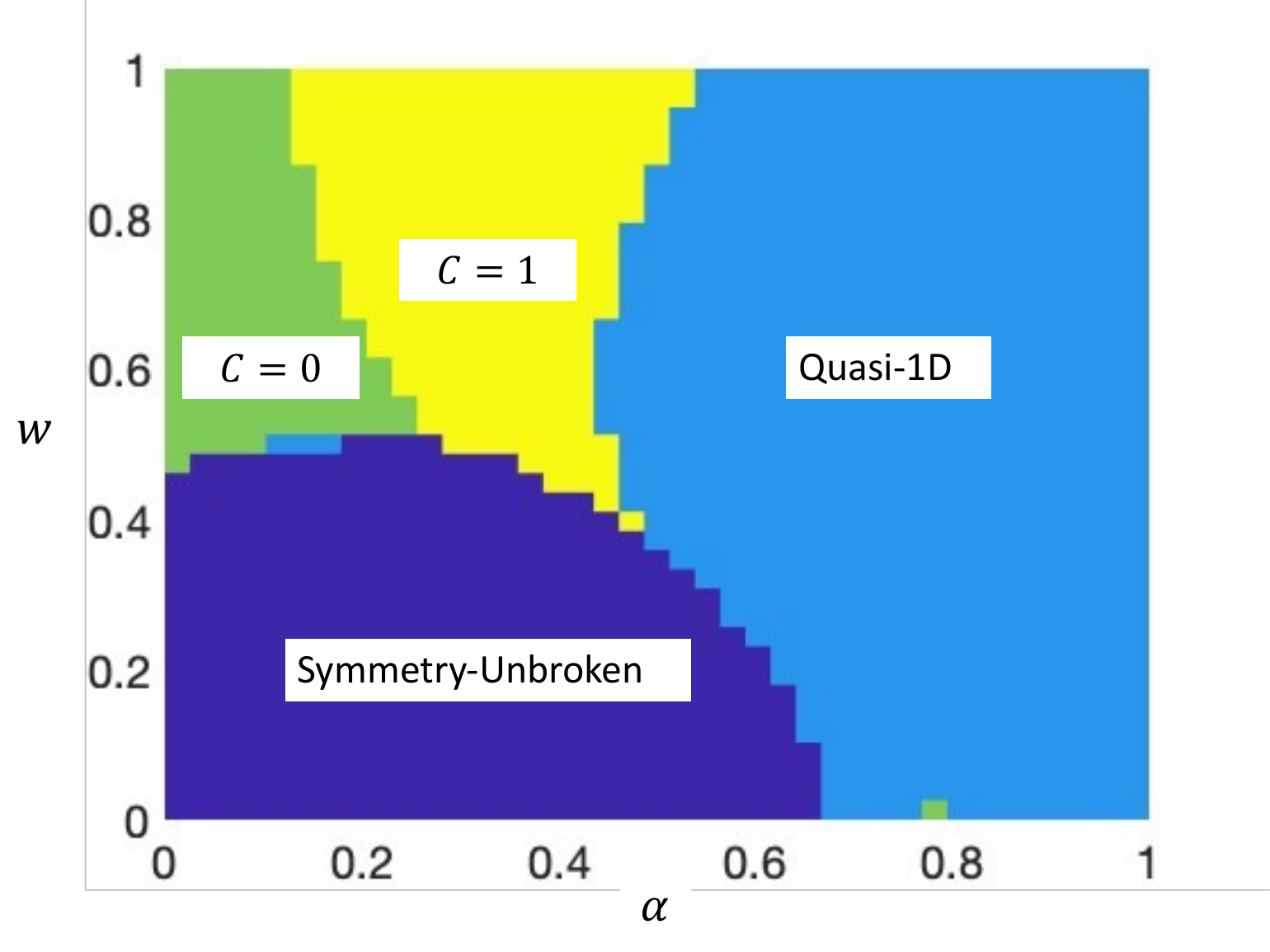}
\caption{Mean-field phase diagram of Eq. ~\eqref{eq:TopologicalToyModel}. Dark blue corresponds to the symmetry-unbroken phase, light blue corresponds to a quasi-1D phase where only one component $\Delta_a$ condenses, green corresponds to the topologically trivial phase where both dipole components condense, and yellow corresponds to a Chern number of $+1$. The calculation is done with parameters $h=0.5$, $r_0^{(x)}= r_0^{(y)} = 0.8$.}
    \label{fig:DipolarTopoPhaseDiagram_Toy}
\end{figure}

We perform a restricted mean-field theory for the order parameter $\Delta^a$ using functional integral\footnote{There is a normal ordering step that we overlooked when passing Eq. \ref{eq:TopologicalToyModel} to a path integral; however if $\mathcal{A}^a_{ii} = 0$, the normal ordering does not generate any new terms.}. After a Hubbard-Stratonovich transform, the action for the order parameter and the fermions takes the form

\begin{equation} \label{eq:HSPartitionFunction}
\begin{split}
\mathcal{S}&=  \int d\tau\biggr(\int d^2x\, \frac12r^{(a)}_0\abs{\Delta_a(x)}^2+\frac12K_{ab}\abs{\partial_a\Delta^b}^2 \\
&+\sum_{i} \biggr(\sum_a(\Delta^*_{i,a}\hat{D}^a_i + \Delta_{i,a}(\hat{D}^a_i)^\dag)+\Psi^\dag_i(\partial_\tau-\mu + M)\Psi_i  \biggr)\biggr).
\end{split}
\end{equation}
The mass term $\frac12r_0$ is given by the zero-momentum component of $-(\mathcal{A}^a)^{-1}_{ij}$. We assume that $\mathcal{A}^a$ is such that $r_0>0$; if this is not the case, then the $\mathcal{A}^a_{ij}$ expansion would instead need to be carried out around a momentum of $(0,\pi)$ or $(\pi,0)$ to ensure a stable mean-field theory. Now, we seek the energy functional for the static, uniform $\Delta_a$ field. This is easy to do since the second line of Eq. \ref{eq:HSPartitionFunction} describes a two-band Hamiltonian at half-filling, and we can easily integrate out the fermions in this case to arrive at the energy functional (for constant order parameter)
\begin{equation}\label{eq:Energy_Functional}
F(\Delta_a)= \frac12r^{(a)}_0\abs{\Delta_a(x)}^2+ E(\Delta_a)
\end{equation}
where $E(\Delta_a) = \int\frac{d^2k}{(2\pi)^2}\xi_-(\vec{k})$ is the energy of the occupied band of the mean-field Hamiltonian. To gain some intuition for the mean-field theory, we expand the energy functional to quadratic order:
\begin{equation} \label{eq:energy_functional_quadratic}
\begin{split}
F^{(2)}(\Delta_a) &= \int\frac{d^2k}{(2\pi)^2}\biggr(\biggr(\frac12 r_0^{(x)} - \frac{\tr(\tilde{T}_x^2)}{4h} + \frac{(\tr(\tilde{T}_x\sigma_z))^2}{8h} \biggr)\Delta_x^2\\
&+\biggr(\frac12 r_0^{(y)} -\frac{\tr(\tilde{T}_y^2)}{4h} + \frac{(\tr(\tilde{T}_y\sigma_z))^2}{8h} \biggr)\Delta_y^2\biggr)
\end{split}
\end{equation}
where $\tilde{T}_a = T_ae^{ik_a}+h.c.$. Roughly, component $\Delta_a$ orders when the coefficient of $\Delta_a^2$ becomes negative. We can see that $\Delta_a$ orders when the dipole hopping (tuned by $w,\alpha$ in Fig. \ref{fig:DipolarTopoPhaseDiagram_Toy}) overcomes the magnetization $M$ (tuned by $h$), which favors a featureless spin-polarized state without symmetry breaking. 

Given the previous choice of matrices $T_x, T_y, M$, we compute the mean-field phase diagram as a function of the parameters $w$, $\alpha$. The results are shown in Fig. \ref{fig:DipolarTopoPhaseDiagram_Toy}. Interestingly, we observed phases with dipolar symmetry breaking and nontrivial topological band structure signatured by non-zero Chern number. The ordering transitions are all first order; this is a common characteristic of dipole ordering transitions, as seen in the mean-field analysis of the bosonic system studied in \cite{LakeDBHM} and compressible fermionic systems in \cite{lakeDFHM}. The topological phase transition at the mean-field level is described by a massless Dirac fermion. This transition is of particular interest because the nature of this topological transition is strongly influenced by the presence of Goldstone modes of the broken dipolar symmetry, and this will be discussed in details in Sec. \ref{Sec:QHTransition}. 

The purpose of this toy model is to illustrate the existence of possible topological phases and phase transitions in dipole condensed phases. The choice of matrices $T_x, T_y$ in Eq. ~\eqref{eq:Choice_of_Matrices} is by design. In real systems, possible Hubbard-like interactions\cite{LakeDBHM,lakeDFHM} may complicate the phase diagram. An interesting question is how we can approximate this model by using tilted optical lattices.  In App. \ref{App:Tilted}, we discuss a tilted optical lattices implementation of the designed model. Compared to Eq. \ref{eq:effectiveTiltHamiltonianPlatonicForm}-\ref{eq:effectiveTiltHamiltonianPlatonicForm_ONSITE_QUARTIC_NORMORD_SIMPLE}, Hubbard-like 4-fermion interaction terms do emerge naturally. These terms could drive a Mott-insulating phase for smaller values of $\Delta_a$, which will ultimately change the nature of the mean-field energy landscape and qualitatively change the phase diagram in Fig. \ref{fig:DipolarTopoPhaseDiagram_Toy}. However, we expect that the universal physics discussed in the rest of this work, particularly the topological phases and phase transitions, will not depend on such microscopic details.

\subsection{Effects of topology on the Goldstone modes}
\label{Sec:Goldstone_EFT}

Having discussed the mean-field phase diagram, we now add the most relevant low energy fluctuations which are the Goldstone modes of the broken dipolar symmetry. We suppose that both components of the dipole symmetry are spontaneously broken. The quadratic part of Goldstone action can be derived from Eq.~\eqref{eq:HSPartitionFunction}. The coefficients will depend on the dipole hopping matrix $\mathcal{A}^{a}_{ij}$. As discussed in \cite{anakru2023nonfermi}, the low-momentum Goldstone-fermion interaction is given by coupling the two Goldstones $\varphi^{x},\varphi^y$ to the quadratic mean-field action in the same fashion as the spatial part of a $U(1)$ gauge field. Despite the fact that the Goldstones are coupling to the fermions like a spatial gauge field, the theory has no gauge symmetry; in the case of Ref. \cite{anakru2023nonfermi}, this lack of gauge symmetry manifested as an additional longitudinal mode that was physical but did not relevantly couple to the fermions in the low energy. Thus, we refer to the Goldstone modes of the dipole symmetry as a "pseudo-gauge field". 

When the dipole condensed phase forms a mean-field band insulator, the bulk fermions are gapped, we may integrate the fermions out to get a Goldstone effective field theory:
\begin{equation}\label{eq:Effective_Goldstone_Bare}
   \mathcal{L}_{eff}[\varphi^a] = -\frac12K_{ab}(\partial_a\varphi^b)^2 +\Delta\mathcal{L}
\end{equation}
where the first term comes from the low-momentum expansion of the dipole-hopping matrix $\mathcal{A}^{a}_{ij}$ as discussed previously, and the $\Delta\mathcal{L}$ comes from integrating out the fermions. Because the Goldstones couple like a $U(1)$ gauge field, determining the quadratic part of $\Delta\mathcal{L}$ is equivalent to computing the linear electromagnetic response of the mean-field band insulator. Thus, we propose the most general quadratic Goldstone action (up to $O((\omega,p)^2)$) for a band insulator (in real time):
\begin{equation}\label{eq:Effective_Goldstone}
   \mathcal{L}_{eff}[\varphi^a] =\frac12\kappa^{ab} (\partial_t\varphi^a)(\partial_t\varphi^b) -\frac12\tilde{K}_{ab}(\partial_a\varphi^b)^2 - \frac{C}{4\pi}\varphi^a\partial_t\varphi^b\epsilon_{ab}
\end{equation}
The $\kappa^{ab}$ term and $\tilde{K}_{ab}$ describe generic renormalized dispersion of the Goldstone modes. The unusual feature of the Goldstone action is the term proportional to $C$ where $C$ is the Chern number of the filled band. This term is an analog of a Chern-Simons term for the purely spatial pseudo-gauge field $\varphi^a$. This term is linear in $\partial_t$ and can be interpreted as a "Berry phase" term -- as explained in App. \ref{App:Quantization}, the mean-field ground state has a nontrivial Berry phase in the space of global dipole symmetry transformations. 

The important effect of the Chern-Simons term is to change the dispersion relation of the Goldstone modes. If $C\neq 0$, one of the Goldstones becomes \textit{quadratically} dispersing (also known as a type-II Goldstone mode\cite{nielsen_chadha}) at low momenta while the other becomes gapped. The total number of gapless Goldstones is less than the number of broken symmetry generators. This effect is reminiscent of the quadratic dispersion of the spin  waves in an $SO(3)$-symmetric Heisenberg ferromagnet, and in fact can be understood in a similar framework as explained below.

Though the effective Lagrangian Eq. \ref{eq:Effective_Goldstone} can be found by straightforward computation of the fermion determinant, it is useful to view it in light of the general considerations of Watanabe and Murayama in Ref. \cite{Watanabe_Goldstone_Theorem}. Here, the coefficient of any term $\varphi^a\partial_t\varphi^b\epsilon_{ab}$ in a Goldstone theory is proportional to the ground state expectation $\avg{[\hat{Q}_a,\hat{Q}_b]}$, $\hat{Q}_a$ being the generators of the broken symmetry. However, the microscopic dipole symmetry generators (which, in a single-particle picture, are just the particle position operators $\hat{x},\hat{y}$) commute, so the relation to the aforementioned result is not immediately obvious. To resolve this, we note that the result of Ref. \cite{Watanabe_Goldstone_Theorem} implicitly assumes that all non-Goldstone degrees of freedom have been integrated out—in particular, this includes the gapped fermionic excitations of the mean-field band insulator. Integrating out such excitations is tantamount to projecting to the occupied band(s). Then, if the occupied bands have a nonzero Chern number, the \textit{projected} single-particle dipole symmetry operators (i.e. the projected \textit{position} operators $\mathcal{P}{x}^a\mathcal{P}$) have a nonzero commutator proportional to the Chern number \cite{resta_chern_marker}, and then the origin of the linear-in-$\partial_t$ term follows from Ref. \cite{Watanabe_Goldstone_Theorem}.  

This point of view also helps to understand that the coefficient of the term $\varphi^a\partial_t\varphi^b\epsilon_{ab}$ remains quantized even after renormalization, since the coefficient is related to the projective representation of the position operators, which is quantized in a gapped system without ground state degeneracy. In App. \ref{App:Quantization}, we discuss a proof of this quantization in the context of our particular setup; physically, the reason behind the quantization is identical to the reason that uniquely gapped systems have a quantized Hall conductance. Thus, the Goldstone action is seen to probe the bulk topology (equivalently, the noncommutative geometry) of the mean-field ground state -- the Chern number of the mean-field ground state is nonzero if and only if there is a single Goldstone with quadratic dispersion as opposed to two Goldstones with linear dispersions. The actual value of the Chern number is difficult to detect just by studying the Goldstone spectrum. However, there is a mode of the Goldstone fields that is gapped due to the linear-in-$\partial_t$ term. This gap depends on the Chern number $C$ but also on the non-universal parameters such as $\kappa^{ab}$. 

We note that a theory similar in structure to Eq.~\eqref{eq:Effective_Goldstone} appears in \cite{Watanabe_Murayama_Redundancies_PRL,nguyen_moroz_melting_vortex_crystal_lifshitz,moroz_hoyos_tkachenko,yi_hsien_Tkachenko,brauner2024dipole} as an effective theory of a rotating vortex lattice in a superfluid which hosts an emergent noncommutative dipole symmetry. The rotating vortex lattice can be parametrized with degrees of freedom in a noncommutative spacetime (i.e. the lowest Landau level), upon which the magnetic translations act on the superfluid mode as a noncommutative dipole symmetry, leading to a quadratically dispersing "Tkachenko mode".  In the present setting of a dipole-conserving Chern insulator, one may also interpret the degrees of freedom as living in a noncommutative spacetime, but due to the underlying lattice this noncommutativity is quantized in contrast to the continuously varying magnetic length in the vortex lattice case.

\section{2-D Topological Phase Transition in the Presence of Goldstone Modes}
\label{Sec:QHTransition}
Now we consider that parameters are tuned such that the Chern number of the mean-field band structure changes from a topological nontrivial band to a trivial one. At the phase boundary, the mean-field theory can be described by a Dirac fermion in (2+1)-D. Indeed, the continuum limit of our engineered mean-field ground state from Eqs. \ref{eq:QuadraticMeanField}-\ref{eq:Choice_of_Matrices} realizes a Dirac fermion at the topological phase transition. Including the Goldstone fluctuation amounts to couple the Dirac fermion to the pseudo-gauge fields. The effective field theory can be written as
\begin{equation}\label{eq:QHTransition_Micro}
\begin{split}
    \mathcal{L} &= \psi^\dag\gamma^0(\fsh{\partial}-\fsh{\varphi}+iM)\psi -\frac12\frac{1}{4\pi}i\epsilon^{ab}\varphi^a\partial_\tau \varphi^b\\
    &+ \frac{1}{2g^2}\biggr(\sum_a(\partial_\tau\varphi^a)^2+v_1^2\sum_{ab}(\partial_a\varphi^b)^2
    +v_2^2(\sum_a\partial_a\varphi^a)^2\biggr).
    \end{split}
\end{equation}
The gamma matrices $\gamma^0=\sigma^1,\gamma^1=\sigma^2,\gamma^2=\sigma^3$ furnish a representation of the Clifford algebra $\{\gamma^\mu,\gamma^\nu\}=2\delta^{\mu\nu}$, $\fsh{A}=A_\mu\gamma^\mu$, $\varphi_\mu = (0,\varphi^a)$, and the Fermi velocity of the Dirac cone has been set to unity. The inclusion of the half-quantized Berry phase term is necessary to ensure that Eq. \eqref{eq:QHTransition_Micro} reproduces Eq. \eqref{eq:Effective_Goldstone} for $C=0,1$ (depending on the sign of $M$) once the massive Dirac fermion is integrated out. Such a term can be justified by manually integrating out the so-called spectator fermions occupying the Brillouin zone (excluding a neighborhood of the Dirac point) \cite{Bernevig_Book_2013}, and it is exactly analogous to the usual manifestation of the parity anomaly of (2+1)-D Dirac fermions when coupled to a gauge field\cite{burkov_half_quantized_parity_anomaly}. 

For analytical control, we study a generalized Lagrangian involving $N$ identical flavors of Dirac fermions, namely
\begin{equation}\label{eq:QHTransition_Micro}
\begin{split}
    \mathcal{L} &= \sum_{i=1}^N\psi^\dag_i\gamma^0(\fsh{\partial}-\frac{ig}{\sqrt{N}}\fsh{\varphi}+iM)\psi_i -\frac12\frac{g^2}{4\pi}i\epsilon^{ab}\varphi^a\partial_\tau \varphi^b\\
    &+ \frac{1}{2}\biggr(\sum_a(\partial_\tau\varphi^a)^2+v_1^2\sum_{ab}(\partial_a\varphi^b)^2
    +v_2^2(\sum_a\partial_a\varphi^a)^2\biggr)
    \end{split}
\end{equation}
We have also rescaled the boson field such that the coupling constant $g$ appears in the fermion-boson coupling. For the purposes of a large-$N$ expansion, we consider $g^2 = O(1)$, so that the RPA diagram (Fig. \ref{fig:RPA_Bubble_Bare}) is the only self-energy contribution to the leading order in the large-$N$ limit. We have taken both the Goldstone and Dirac actions to be rotationally symmetric for conceptual simplicity, and we breifly discuss the role of anisotropy in the end. 

This Lagrangian has a striking resemblance to the famous theory of Dirac fermions in $(2+1)$-d coupled to a gauge field with a half-quantized Chern-Simons term, as studied, for instance, in Ref. \cite{mpa_fisher_mott_transition_anyon_gas}, which is relevant for topological phase transitions in quantum Hall systems\cite{jongyeon_QED3_FCI,ruochen_yinchen_QED3_FCI}. However, the lack of gauge invariance of the Goldstone bosons actually leads to properties that are fundamentally different from those of a gauge theory.

\subsection{RPA Propagators}
We first concern ourselves with the computation of the RPA Goldstone propagator $D_{ab}(\omega,\mathbf{p})$. Because the interaction $-i\psi^\dag\gamma^0\fsh{\varphi}\psi$ is relevant in $(2+1)$-d, the RPA Goldstone self-energy (Fig. \ref{fig:RPA_Bubble_Bare}) is a highly relevant correction compared to the bare kinetic term in Eq. \ref{eq:QHTransition_Micro}. As discussed in greater detail in App. \ref{App:RPAGoldstone}, the RPA propagator exhibits two singularities with $\omega\sim p^{3/2}$ and $\omega\sim p$ at low momentum. We suppose that the pole $\omega\sim p^{3/2}$ eventually leads to a dynamical exponent $z=3/2$ and use this new scaling to systematically discard terms in the RPA propagator that are irrelevant or nonsingular. Then, we may decompose the important contributions to the RPA propagator in terms of a transverse and longitudinal pole:

\begin{figure}
    \centering
    \includegraphics[width=0.4\linewidth]{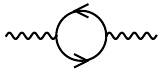}
    \caption{1-loop self-energy diagram for the Goldstone boson}
    \label{fig:RPA_Bubble_Bare}
\end{figure}

\begin{equation}\label{eq:RPA_propagator_relevant_main}
    \begin{split}
        D_{ab}(\omega,\mathbf{p}) &= D^L_{ab}(\omega,\mathbf{p})+ D^T_{ab}(\omega,\mathbf{p})\\
        D^L_{ab} & \sim \lambda^2\frac{p}{ \omega^2+\zeta^2p^3 }\mathcal{P}^L_{ab} \\ 
        D^T_{ab} &\sim \lambda^2\frac{1}{\sqrt{\omega^2+p^2}} \mathcal{P}^T_{ab}
    \end{split}
\end{equation}
where $\lambda^2 = \frac{8}{g^2(1+\pi^{-2})}$ and $\zeta = \lambda v_L$. Here, $\mathcal{P}_{ab}^L = \delta_{ab}-\mathcal{P}_{ab}^T = p_ap_b/p^2 $ is the projector into the longitudinal component $\varphi_L$ of the Goldstone modes, i.e. the component of the fields $\varphi^a(\vec{p})$ parallel to the momentum $\vec{p}$. $\mathcal{P}_{ab}^T$ projects to the transverse component $\varphi_T$ with $p_a\varphi^a_T = 0$. Accordingly, we see that the interaction with the Dirac fermions generates a singular self-energy $\sim \omega^2/p$ for the longitudinal Goldstone. A Goldstone mode with a similar fractional-power dispersion $\omega\sim p^{3/2}$ appeared in \cite{watanabe_murayama_ripplon_32} as a 2-D interfacial mode; it will be interesting to investigate any relationship between these two settings. The transverse Goldstone mode also receives a self-energy that dominates its bare kinetic term and retains a linear dispersion. 

\begin{figure}
    \centering
\includegraphics[width=0.4\linewidth]{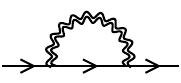}
    \caption{1-loop self-energy diagram for the fermions. Only the longitudinal piece $D_{ab}^L$ of the fully corrected RPA propagator of the Goldstone (Eq. \ref{eq:RPA_propagator_relevant_main}) is used and indicated by the double wavy line.}
    \label{fig:RPA_Boson_Fermion_SE}
\end{figure}

Now we must concern ourselves with the fermion self-energy. By the form of the RPA propagator in Eq. \ref{eq:RPA_propagator_relevant_main}, we see that the $\varphi_T$ field has a higher scaling dimension than the $\varphi_L$ field, so we neglect the transverse part of the RPA propagator in all computations. We will see later that once the scaling dimensions have been fully determined for all fields in the effective action, the interaction with the transverse Goldstone $\varphi_T$ interacts with the fermions irrelevantly. In App. \ref{App:OneLoopFermionSE}, we evaluate the usual one-loop diagram for the fermion self-energy (Fig. \ref{fig:RPA_Boson_Fermion_SE}), finding a contribution
\begin{equation}\label{eq:Bare_Fermion_Self_Energy}
\Sigma^{(1)}(\nu,\mathbf{q}) =  \frac{\lambda^2 g^2}{4\sqrt{2}\zeta N}\gamma^0\abs{\nu}^{1/2}\sign(\nu) + O(\nu,\abs{q}).
\end{equation}
This tells us that the bare Dirac action is not simply renormalized as in the case of $\mathrm{QED}_3$ or the quantum Hall transition discussed in \cite{mpa_fisher_mott_transition_anyon_gas}, but in fact replaced with more relevant contributions with a non-analytic behaviour as $\omega,\vec{q}\to 0$, due to the soft longitudinal mode which is absent in the gauge theory. 

\subsection{Self-Consistent Self-Energies}
\label{SubSec:SC_Fermion_Self_Energy}

The complicated structure of the singular self-energies for the bosons and fermions poses a difficulty in evaluating higher loop corrections. However, in the limit where the number of Dirac fermions $N\gg 1$, diagrams may be organized in terms of their powers of the small quantity $N^{-1}$; some of these diagrams may be further organized into Dyson equations for the fermion and Goldstone boson propagators, with the rest contributing only subleading terms in the large-$N$ expansion. The process of determining the fermion and boson self-energies can be formulated in terms of a self-consistent cycle, where the RPA Goldstone boson is used to generate a singular self-energy for the fermion, which then feeds back to the boson self-energy. We terminate this cycle once corrections to the self-consistent solution become subleading in $N^{-1}$.

To begin this self-consistency cycle, we first focus on the possibility of further non-analytic corrections to the fermion propagator. To this end, we propose an ansatz for the self-energy of the Dirac fermion: 
\begin{equation}\label{eq:Fermion_SE_Ansatz}
\Sigma(\nu,\vec{q}) = Z_0\abs{\nu}^{\alpha}\sign(\nu)\gamma^0+Z_1\abs{q}^\beta\gamma_{\vec{q}} + \ldots
\end{equation}
where $\gamma_{\vec{q}} = \vec{\gamma}\cdot\hat{q}$. Such a self-energy respects the assumed rotational symmetry and should be consistent with the dynamical exponent $z=3/2$ derived from the RPA Goldstone propagator. Furthermore, for $\alpha,\beta<1$, this will render the bare part of the Dirac action irrelevant, so the fermion propagator in the low energy can be taken to be $(\Sigma(\omega,\vec{q}))^{-1}$. Using this ansatz, we may self-consistently fix the fermion propagator by demanding that it is consistent with the one-loop diagram for the self-energy (i.e., that it solves a Dyson equation as shown in Fig. \ref{fig:Self_consistent_fermion}). This is equivalent to the resummation of the so-called rainbow diagrams, a strategy used in (e.g.) \cite{torroba_thermal_NFL}. The one-loop diagram for the fermion self-energy is evaluated using the Fermion propagator $(\Sigma(\omega,\vec{q}))^{-1}$:

\begin{figure}
    \centering
\includegraphics[width=0.8\linewidth]{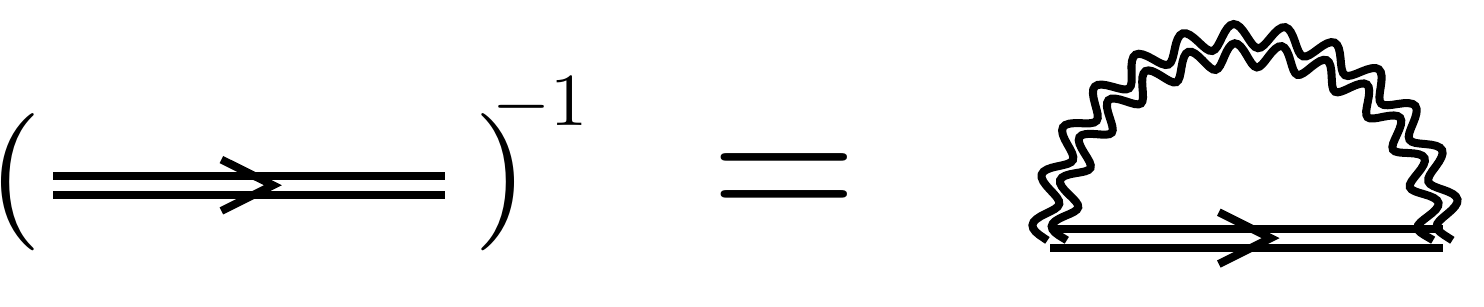}
    \caption{Self-consistency equation for the fermion self-energy $\Sigma(\nu,\vec{q})$, given by Eqs. \eqref{eq:Fermion_SE_Ansatz}-\eqref{eq:DiracSelfConsistent_Undetermined}. Double wavy line represents the longitudinal piece $D_{ab}^L$ of the RPA propagator in Eq. ~\eqref{eq:RPA_propagator_relevant_main} and the double fermion line corresponds to the self-consistently determined fermion propagator, which is given by $(\Sigma(\omega,\vec{q}))^{-1}$ in the IR limit.}
    \label{fig:Self_consistent_fermion}
\end{figure}

\begin{equation}\label{eq:DiracSelfConsistent_Undetermined}
\begin{split}
    \Sigma(\nu,\vec{q}) &= -{\lambda^2}\int d\omega d^2p \frac{p_ap_b/p}{ \omega^2+\zeta^2\mathbf{p}^3 } \\
&\times \frac{\gamma^a(Z_0\abs{\omega+\nu}^{\alpha}\sign(\omega+\nu)\gamma^0+Z_1\abs{p+q}^\beta\gamma_{\vec{p}+\vec{q}})\gamma^b}{Z_0^2\abs{\omega+\nu}^{2\alpha}+Z_1^2\abs{p+q}^{2\beta}}
    \end{split}
\end{equation}

\begin{figure}
    \centering
\includegraphics[width=0.4\linewidth]{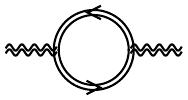}
    \caption{Correction to the boson propagator using the fermion propagator found in Eq. \ref{eq:Self_energy_solution_main} (represented by straight double lines).}
    \label{fig:boson_stronger_bubble}
\end{figure}

A simple dimensional analysis using the relation $[\omega]=\frac{3}{2}[p]$ reveals $\alpha=\frac12$, $\beta=\frac34$ which is indeed consistent with the $z=3/2$ scaling. In App. \ref{App:FermionSCSE}, we justify the scaling form of the self-energy more explicitly and show that

\begin{equation}
    \label{eq:Self_energy_solution_main}
    \Sigma(\nu,\vec{q}) = \frac{\lambda g}{\zeta\sqrt{N}}\biggr(c_1\abs{\nu}^{1/2}\sign(\nu)\gamma^0 +c_2\zeta^{1/2}\abs{q}^{3/4}\gamma_{\vec{q}} \biggr) + \ldots.
\end{equation}
where $c_1,c_2$ are numerical constants that satisfy a complicated transcendental equation. Notice, the $N$-scaling of the self-energy is corrected to be $O(N^{-1/2})$ by the self-consistency procedure compared to the one-loop result. 
As discussed in App. \ref{App:FermionSCSE}, the boson propagator must also be fixed self-consistently, which is tantamount to including certain self-energy diagrams beyond RPA; specifically, these are the so-called bubble diagrams that incorporate the self-consistent fermion propagator from Eq. \eqref{eq:Self_energy_solution_main} (see Fig. \ref{fig:boson_stronger_bubble}). This generates a more complicated term $N\frac{\zeta^2}{\lambda^2} p^2 F(\omega/\zeta p^{3/2})$ which, in the large-$N$ limit, dominates the RPA kinetic term. Nevertheless, the new boson propagator preserves $z=3/2$. At this point, the self-consistency cycle can be terminated because any further corrections to the fermion kinetic term using this new boson propagator are suppressed by a factor of $1/N$ compared to Eq. \ref{eq:Self_energy_solution_main} -- importantly, any such corrections simply renormalize the coefficients in Eq. \ref{eq:Self_energy_solution_main} instead of generating more singular terms.

Accordingly, we can write down an effective action involving the self-consistent boson and fermion propagators that reflects the dominant terms in the large-$N$ expansion:
\begin{equation}\label{eq:OneLoopEffectiveAction}
\begin{split}
&\mathcal{S}_{eff} = \int d^2pd\omega\,\biggr(\frac12\varphi_Lp^2F(\abs{\omega}/\zeta p^{3/2})\varphi_L \\
&+ \sum_i\bar{\psi}_i(c_1\abs{\omega}^{1/2}\sign(\omega)\gamma^0+c_2\zeta^{1/2}\abs{p}^{3/4}\gamma_{\vec{p}})\psi_i\biggr)   \\
&- \frac{i\lambda}{\sqrt{N}} \int dtd^2x \sum_i\bar{\psi}_i\fsh{\varphi}_L\psi_i \\
&+ \int d^2pd\omega\,\frac12\varphi_T(\omega^2+p^2)^{1/2}\varphi_T
\\
&- i\zeta\int dtd^2x \sum_i\bar{\psi}_i\fsh{\varphi}_T\psi_i.
    \end{split}
\end{equation}
In writing the above action, we have performed the rescalings $\varphi_L\to N^{-1/2}\frac{\lambda}{\zeta}\varphi_L$, $\psi_i\to (\zeta\sqrt{N}/\lambda g)^{1/2}\psi_i$, and $\varphi_T\to \lambda \varphi_T$. Note that we have scaled the fields $\varphi_T$, $\varphi_L$ by different factors on account of their different scaling dimensions. Under the tree-level $[\omega]=3/2[p]$ scaling that keeps the above interaction-driven kinetic terms invariant\footnote{Note that under this scaling, $\sqrt{\omega^2+p^2}\to \abs{p}$.}, we see that the interaction strength $\lambda$ between the fermions and the longitudinal Goldstone is actually marginal, whereas in the bare action this interaction was relevant. In addition, the interaction $\bar{\psi}_i\fsh{\varphi}_T\psi_i$ with the transverse Goldstone is \textit{irrelevant}; this justifies our neglect of the this mode throughout the computations. 

We conjecture that this theory of a Dirac fermion with a single strongly-coupled collective mode, along with another decoupled collective mode, describes a novel interacting fixed point. Within the scope of the diagrammatic calculations described by the above action, we predict a dynamical exponent $z=3/2$ as described above. Given that Eq. \ref{eq:OneLoopEffectiveAction} was self-consistently fixed by only the "bubble" and "rainbow" diagrams, there is the possibility of higher-loop corrections to the $z=3/2$ result. However, if we take the limit $N\gg 1$, loop corrections to the fermion-boson vertex are suppressed by a factor of $N^{-1}$, suggesting that the tree-level scaling dimensions from this effective action become exact as $N\to\infty$ and that corrections are organized in powers of $N^{-1}$. We leave a detailed study of this $1/N$ expansion to future work. 


In discussing this interacting fixed point, we must also consider the fate of other symmetry-allowed operators that were less relevant than the fermion-Goldstone interaction, in case they become marginal or relevant at the new fixed point. For example, a 4-fermion interaction like $\sigma(\bar{\psi}\psi)^2$ is irrelevant at the Gaussian fixed point, since it has dimension $[\sigma]=-1[p]$. At this new fixed point, $[\sigma]=-2[p]$—i.e., the 4-fermion interaction has become \textit{more} irrelevant at the fixed point, and it is unlikely that corrections in the $1/N$ expansion could cause this scaling dimension to be positive. Another class of symmetry-allowed operators are cubic or higher terms for the Goldstone modes, since such nonlinear terms are always generated in the process of integrating out high-energy fermionic modes; however, these operators always enter with spatial or temporal derivatives since they must preserve the shift symmetry $\varphi_a\to\varphi_a + c_a$. The most relevant operator of this kind we could write down takes the form $(\partial_a \varphi_b)^3$, whose coefficient has a scaling dimension of (at least) $-\frac{3}{2}[p]$. Of course, a fermion mass $M$ is still relevant and serves as the tuning parameter for the transition. A Goldstone mass term $\frac12 M_{ab}\varphi_a\varphi_b$ is also obviously relevant with $[M_{ab}]=2[p]=\frac{4}{3}[\omega]$ at the fixed point. This scaling dimension could govern (e.g.) low-temperature properties in the experimentally relevant case where the dipolar symmetry becomes slightly broken, along the lines of a similar discussion in \cite{anakru2023nonfermi}.



We note that the strategy of choosing a consistent ansatz for the fermion propagator is reminiscent of the interaction-driven scaling that has been used to approach quantum critical theories where the quasiparticle propagators become strongly renormalized by interactions \cite{schlief_lunts_lee_exact_critical_afm,lunts_schlief_lee_emergence_control_afm,sung_sik_NFL_review}. Generally, a bare Lagrangian with a interaction that is relevant under Gaussian scaling dimensions will exhibit singular quasiparticle self-energies; as has been done in the previous discussion, one then demands the singular self-energies to be invariant under the RG scaling, leading to modified field dimensions and dynamical exponent that render the formerly relevant interaction marginal.  

\begin{figure}
    \centering
    \includegraphics[width=\linewidth]{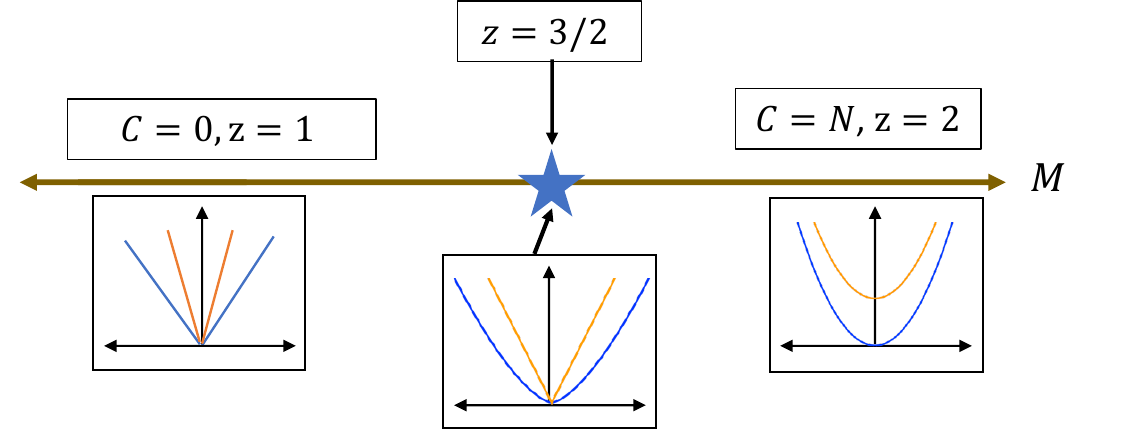}
    \caption{Phase diagram of Eq. \eqref{eq:QHTransition_Micro} tuned by the Dirac mass $M$, with the dynamical exponent and mean-field Chern number of each phase labelled. Dispersion relations of the $\varphi_a$ degrees of freedom are shown: there are two linear dispersing Goldstones for $C=0$, one quadratically dispersing and one gapped Goldstone for $C\neq 0$, and two Goldstones with $\omega\sim p^{3/2}$ and $\omega\sim p$ at the critical point.}
    \label{fig:QH_Phase_Diagram}
\end{figure}

\subsection{Analogies to Gauge Theory}

In models of dipolar symmetry breaking, it is tempting to draw some analogy with the problem of matter coupled to an emergent $U(1)$ gauge field since the dipolar Goldstones couple as a pseudo-gauge field. In a technical sense, no such analogy is ever guaranteed because the Goldstone boson action is not gauge invariant; this absence of gauge invariance manifests in the presence of another longitudinal propagating degree of freedom as compared to the gauge theoretic case\footnote{Another important distinction between the two classes of problems is that there is no pseudo-gauge field analog of the timelike $A^0$ component. Without a Chern-Simons term, the $A^0$ component is traditionally integrated out and its only effect is to generate a Coulomb interaction for the matter.}. In certain cases like the problem of a Fermi surface coupled to dipolar Goldstone modes (as treated in \cite{anakru2023nonfermi}), this distinction is not very important in the IR because the longitudinal Goldstone mode dynamically decouples from the Fermi surface, and the low-energy theory is thus nearly isomorphic to the gauge-theoretic case. However, in the present situation of a (2+1)-D Dirac fermion coupled to dipolar Goldstone modes, it is the longitudinal mode that remains strongly coupled while the transverse mode decouples, and the theory ultimately acquires a non-relativistic scaling with dynamical exponent $z=3/2$. Thus, there is no mapping to QED3 as studied in (e.g.) \cite{mpa_fisher_mott_transition_anyon_gas}, which remains Lorentz-invariant and whose boson propagator is purely transverse. We also note that in the case of QED3 with a Chern-Simons term, the relevant interaction between fermions and the transverse gauge field becomes marginal once the gauge field kinetic term was replaced by the interaction-generated terms; a similar phenomenon occurs in our in our setting, but the presence of a soft longitudinal mode further modifies the scaling dimension of the fermion fields, which drives the coupling to $\varphi_T$ into irrelevancy.

\subsection{Role of Anisotropy}

In our treatment so far, the bare and interaction-generated contributions to the Goldstone kinetic term were diagonal in the transverse/longitudinal basis for the $\varphi_a$-fields; this was due to our assumption of rotational symmetry. Cubic anisotropy, which is an inevitable consequence of realizing this theory on a lattice, will (at the very least) lead to bare kinetic terms of the form $\sum_a f_a(\partial_a\varphi^a)^2$ which not only break the rotational symmetry (and, in the case of the tilted lattice implementation in App. \ref{App:Tilted}, the cubic symmetry), but add coupling terms between the transverse and longitudinal Goldstone components. However, if we treat such coupling terms as perturbations to Eq. \ref{eq:OneLoopEffectiveAction}, they turn out to be irrelevant, so the physical picture of a decoupled transverse mode and a strongly-coupled longitudinal mode remains robust to sufficiently weak anisotropy. 

\section{Discussion and Conclusions}
In this work, we study dipolar-symmetric fermionic models at integer fillings. We focus on the properties of spontaneous dipolar symmetry-breaking phases which host band insulators as the mean-field ground state. We study how the fermion band topology influences the properties of the dipolar Goldstone modes. We also investigate the influence of the Goldstone modes on the topological phase transitions of the mean-field ground states. 

In particular, we constructed various effective field theories for dipolar symmetry breaking by relating terms in the Goldstone effective action to quantized topological responses of the gapped mean-field band insulator. In 1-D, we find that the quantized polarization response $\theta\in\{0,\pi\}$ of the mean-field band structure changes the nature of the KT transitions for the dipolar Goldstone theory, and the critical theory separating the two values of $\theta$ was identified with an exotic Lifshitz ($z=2$) theory featuring (e.g.) dipolar long-range order. In 2-D, we identified that a nonzero Chern number of the mean-field band structure modifies the counting and dispersion relation of the Goldstone modes of the dipolar symmetry breaking. The Dirac fermion describing the topological phase transition between the topological and trivial band insulators gets strongly renormalized because of the presence of the dipolar Goldstone modes. This topological transition, although resembling some features of a quantum Hall transition, was found to exhibit a novel critical behavior where both Goldstone degrees of freedom were strongly renormalized, with one Goldstone mode ultimately leading to a dynamical critical exponent $z=3/2$ and the other Goldstone mode dynamically decoupling from the matter degrees of freedom.

Our results shed light on how topological invariants can manifest in gapless phases of matter, and what kinds of critical points can separate such gapless phases. Since the phases and critical behaviors we describe are universal by such topological considerations (and, in the 2-D case, detectable via the dynamical exponent which affects the structure of the tower of states at finite system size), it may be possible to detect these phases and phase transitions numerically. We leave the design and the numerical study of such models for future work. Tilted lattice schemes may also provide an experimental platform for observing the physics we describe. To find a physically feasible tilted optical lattice implementation of the physics is another interesting future direction. Finally, it will eventually be interesting to generalize this investigation to general multipole groups and other topological invariants, potentially drawing connections to more exotic multipolar Chern-Simons theories  \cite{huang2023chernsimons,May-Mann_Hughes_Topo_Dipole_Cons}. We leave such investigations to future work.

\section*{Acknowledgements}
We thank Ribhu Kaul, Jainendra Jain, Ethan Lake, Xiao-Yang Huang, Salvatore Pace, and Arkya Chatterjee for helpful discussions. AA and ZB are supported by startup funds from the Pennsylvania State University.
\bibliographystyle{unsrt} 
\bibliography{arxivsubmission}

\appendix
\onecolumngrid

\section{Exact Edge States in 1-D Dipole-Conserving Models}
\label{App:Micro1D}

In the main text, the presence of edge states themselves was justified within a mean-field framework—i.e., by appealing to the fact that the SSH model regularized on a finite chain has edge modes. It will be interesting to verify the existence of these edge modes using a more exact numerical method; however, there is a simple physical intuition for why such edge modes ought to exist depending on the nature of the dipolar order. 

We refer to the Hamiltonian Eq. \ref{eq:App_Micro_Model_SSH}, which we argued is described by the $\theta=\pi$ phase of Eq. \ref{eq:Effective_Goldstone_SSH} at mean-field level. In the extreme limit where $h=0$, the Hamiltonian is exactly solvable in a sense—i.e., we may find the exact ground state wavefunction and low-lying excitations. The essential feature that we will use for this is that the Hamiltonian \ref{eq:App_Micro_Model_SSH} with $h=0$ preserves the following "Krylov subspace" of the total Hilbert space: 
\begin{equation}
\label{eq:AppKrylovSubspace}
\mathcal{K} = \mathrm{span}_{\{\hat{G}_i\}}\biggr(\prod_{i=1}^{N-1}\hat{G}_i\ket{0}\biggr)
\end{equation}
where $\ket{0}$ is the free fermion vacuum and $\hat{G}_i$ can be either $d^\dag_i$ or $c^\dag_{i+1}$. We can map such states to a length-$(N-1)$ spin chain by letting the choice of fermion occupancies $\{\hat{G}_i\}$ define the spin configuration and equating $d^\dag_i \implies \uparrow$, $c^\dag_{i+1} \implies \downarrow$. Then, as is discussed in \cite{moudgalya_prem_bernevig_krylov_dipole}\footnote{See also \cite{lakeDFHM} for analogous discussion of a similar model.}, the dipole-hopping term simply becomes $-\frac{t_d}{2}(X_iX_{i+1}+Y_{i}Y_{i+1})$ where $X,Y$ are Pauli operators in this new spin basis; i.e., the Hamiltonian becomes an isotropic $XY$-spin chain in this Krylov subspace. The dipolar order parameter $e^{i\varphi_x}$ is mapped to $(X_i+iY_i)$ which has power-law correlations \cite{franchini2017introduction}, and this XY-spin chain is indeed described\footnote{One may prove this using a spin coherent state path integral \cite{fradkin_2013}.} by the $\theta=\pi$ phase of Eq. \ref{eq:Effective_Goldstone_SSH}. Besides these emergent spin degrees of freedom, there were also two fermions $c^\dag_1$ and $d^\dag_N$ from either end of the fermion chain that do not couple to the emergent spins; at half-filling, exactly one of these degenerate orbitals is occupied. This edge degeneracy resembles precisely that of the free SSH chain in the topological phase, except that it coexists with a gapless bulk. We emphasize that this edge degeneracy is clearly a many-body effect and relies on no mean-field approximations. 

\section{Tilted Lattice Implementation}
\label{App:Tilted}
In this section, we motivate the interacting Hamiltonian in Eq. \ref{eq:TopologicalToyModel} by designing a tilted lattice protocol applied to an ordinary tight-binding model. To this end, we introduce a fairly general model of a Hamiltonian in d-dimensions with 2 fermion species (i.e. spinful fermions):
\begin{equation}\label{eq:Tilt_Microscopic_App}
\begin{split}
\hat{H} = \sum_{i,a}\biggr(\Psi^\dag_i M_0 \Psi_i + \Psi^\dag_i T_a\Psi_{i+a} &+ \Psi^\dag_{i+a} T_a^\dag\Psi_{i}\biggr) + \hat{V}_{os}+\hat{V}_{nn} + \sum_aV_a\hat{Q}^a_{dip}\\
\hat{V}_{os}=u\sum_{i} (\Psi^\dag_i\Psi_i) (\Psi^\dag_{i}\Psi_{i}),\,\,\, &\,\,\,\hat{V}_{nn}=U\sum_{i,a} (\Psi^\dag_i\Psi_i) (\Psi^\dag_{i+a}\Psi_{i+a})
\end{split}
\end{equation}
where $T_a=\vec{t}_a\cdot\vec{\sigma}$. Here, we only consider nearest-neighbor hopping as well as hopping between orbitals within a unit cell. $\Psi^\dag = (c^\dag_\uparrow, c^\dag_\downarrow)$ and the term $V\hat{Q}_{dip}$ proportional to the dipole charge operator implements the linear tilt on the system. In the following simple case, we will assume that $\hat{Q}^a_{dip} = \sum_i (\vec{a}\cdot\vec{r}_i)\Psi^\dag_i \Psi_i$, i.e. fermions within the same unit cell carry no net dipole moment. Passing to a rotating frame, and defining $\Omega_a = V_a\abs{a}$, 
$$e^{it\sum_aV_a\hat{Q}_{dip}^a}\hat{H}e^{-it\sum_aV_a\hat{Q}_{dip}^a} = \sum_{i,a}\biggr(\Psi^\dag_i M_0 \Psi_i + e^{-i\Omega_a t}\Psi^\dag_i t_\mu\sigma^\mu\Psi_{i+a} + e^{i\Omega_a t}\Psi^\dag_{i+a} t_\mu^*\sigma^\mu\Psi_{i}\biggr) +  \hat{V}_{os}+\hat{V}_{nn} $$
Following \cite{lakeDFHM}, we can treat this tilted-lattice problem as a (quasiperiodic) Floquet problem and compute the high-frequency effective Hamiltonian, whose formal expression up to 2nd order in the inverse driving frequency $\Omega^{-1}$ is
\begin{equation}\label{eq:HFE2ndorder}
    \hat{H}_{eff} =  -\mu\hat{N} +\mathcal{D} + \sum_{m>0}\frac{1}{m\Omega}[\hat{H}_m,\hat{H}_{-m}] + \sum_{m\neq 0}\frac{1}{2(m\Omega)^2}[\hat{H}_{-m},[\mathcal{D},\hat{H}_{m}]] +\sum_{m\neq 0}\sum_{m'\neq 0, m'\neq m}\frac{1}{3mm'\Omega^2}[\hat{H}_{-m'},[\hat{H}_{m'-m},\hat{H}_{m}]]
\end{equation}
where $\mathcal{D}$ is the dipole-conserving part of the original Hamiltonian (i.e. the time-average of the Floquet Hamiltonian) and $\hat{H}_m$ are the Fourier harmonics. The expression above is valid for a single-frequency drive, but the generalization to the quasiperiodic Floquet problem is straightforward so long as the driving frequencies are incommensurate\footnote{If the drive frequencies are commensurate, i.e.  $\Omega_x/\Omega_y=p/q$, then the problem may be treated as a single drive problem at frequency $q\cdot\min(\Omega_x,\Omega_y)$. However, the effective Hamiltonian will not be dipole-conserving, because it will contain processes that (e.g.) change $\hat{Q}_{dip}^x$ by $p\abs{a}$ and $\hat{Q}_{dip}^y$ by $-q\abs{a}$.}. Computation of the effective Hamiltonian is straightforward but tedious, and we only present the end result:
\begin{equation} \label{eq:effectiveTiltHamiltonianPlatonicForm}
\begin{split}
\hat{H}_{eff} = &-\mu\hat{N} + \sum_i\Psi^\dag_i M \Psi_i +\sum_{ij,a}(\hat{D}^a_i)^\dag\mathcal{A}^a_{ij}\hat{D}^a_j +\hat{H}_{on-site}
\end{split}
\end{equation}
\begin{equation} \label{eq:effectiveTiltHamiltonianPlatonicForm_ONSITE_QUADRATIC}
\begin{split}
M= M_0 + \sum_a\biggr(\biggr(\frac{1}{\Omega_a}+\frac{2u-U}{\Omega_a^2}\biggr)[T_a^\dag,T_a]+\frac{4u-2U}{\Omega_a^2}T_aT_a^\dag+\frac{1}{2\Omega^2_a}([T_a,[M_0,T_a^\dag]]+[T_a^\dag,[M_0,T_a]])\biggr)
\end{split}
\end{equation}
\begin{equation} \label{eq:effectiveTiltHamiltonianPlatonicForm_HOPPINGMATRIX}
\begin{split}
\mathcal{A}^a_{ij} = \frac{1}{\Omega_a^2}\sum_{b}\biggr( &\delta_{ab} (2(U-u)(\delta_{i,j+a}+\delta_{i,j-a})-U(\delta_{i,j+2a}+\delta_{i,j-2a}))\\
+ &(1-\delta_{ab})U(2\delta_{i,j-b}+2\delta_{i,j+b}-\delta_{i,j-a-b}-\delta_{i,j+a+b}-\delta_{i,j-a+b}-\delta_{i,j+a-b}))  \biggr)
\end{split}
\end{equation}
\begin{equation} \label{eq:effectiveTiltHamiltonianPlatonicForm_ONSITE_QUARTIC_NORMORD_SIMPLE}
\begin{split}
\hat{H}_{on-site} =\,&\sum_{i}\biggr(u+\sum_a(2U-4u)\frac{\abs{\vec{t}_a}^2}{\Omega_a^2}\biggr) :n_i n_i:+\sum_{i,a}\biggr(U+\frac{2u+U}{\Omega_a^2}\abs{\vec{t}_a}^2 - \biggr(\sum_b\frac{4U\abs{\vec{t}_b}^2}{\Omega_b^2}\biggr)\biggr) :n_i n_{i+a}:\\
&+\sum_a\frac{2u}{\Omega_a^2}\sum_i:\biggr((\Psi^\dag_{j-a}i(\vec{t}\times\vec{t}^*)\cdot\vec{\sigma}\Psi_{j-a}  - \Psi^\dag_{j+a}i(\vec{t}\times\vec{t}^*)\cdot\vec{\sigma}\Psi_{j+a})(\Psi_j^\dag\Psi_j)\biggr):\\
&+\sum_{ab}\frac{U}{\Omega_a^2}:\biggr(\sum_{i,\pm}(\Psi^\dag_{j-a\pm b}i(\vec{t}\times\vec{t}^*)\cdot\vec{\sigma}\Psi_{j-a\pm b}  -\Psi^\dag_{j+a\pm b}i(\vec{t}\times\vec{t}^*)\cdot\vec{\sigma}\Psi_{j+a\pm b})(\Psi_{j}^\dag\Psi_{j}) \biggr):+\sum_{i,a,b}\frac{2U\abs{\vec{t}_{a}}^2}{\Omega_a^2}:n_i n_{i+a+b}:\\
&+\sum_a\frac{(2u-U)}{\Omega_a^2}:\biggr(-\Psi_{i}^\dag T_a^\dag\Psi_{i} \Psi^{\dag}_{i+a} T_a \Psi_{i+a}-\Psi_{i}^\dag T_a\Psi_{i} \Psi^{\dag}_{i+a} T_a^\dag \Psi_{i+a}  +\abs{\vec{t}_a}^2\vec{S}_i\cdot\vec{S}_{i+a}+i\epsilon^{ij\alpha\beta}(t_{a,i}t_{a,j}^*)S_{i,\alpha} S_{i,\beta}\biggr):
\end{split}
\end{equation}
where $:\hat{A}:$ indicates normal-ordering of the operator $\hat{A}$ and $S_{i,\alpha} = \Psi^\dag_i\sigma_\alpha \Psi_i$, with $\sigma_\alpha = (\mathbbm{1},\vec{\sigma})$ being the 4-vector of Pauli matrices.

The essential difference between this effective Hamiltonian and the simple model studied in the main text is the presence of $\hat{H}_{on-site}$, which is parametrically related to the scale of the dipole-hopping terms and thus cannot be neglected. Furthermore, the last line of $\hat{H}_{onsite}$ provides spin dynamics, which may potentially overpower the spin polarization set by the quadratic term $\sum_i\Psi^\dag_i M \Psi_i$. In general, these neglected terms could open up the possibility of Mott phases or other magnetic orders, and we only expect the mean-field results presented in the main text to hold in a regime where $\hat{H}_{on-site}$ is irrelevant. We further note that anisotropy (and moreover, the breaking of the cubic symmetry), which is neglected in the main text, is essential to the tilted lattice setup; it is a consequence of the tilt strengths $\Omega_x, \Omega_y$ being incommensurate, which is in turn required for $\hat{H}_{eff}$ to be perfectly dipole-conserving at each order in $\Omega_a^{-1}$.
\section{Effective Goldstone Theory and Charge Commutator}
\label{App:CommutatorFromAction}
Here, we will quickly explain how the charge commutator $\avg{[Q_1,Q_2]}$ is related to the linear-in-$\partial_t$ term in the Goldstone effective theory (Eq. \ref{eq:Effective_Goldstone}). It is sufficient to analyze zero-momentum fluctuations, described by the Lagrangian:

\begin{equation}\label{eq:AppendixToyLagrangian}
    L = \frac12\kappa_1\dot{q}_1^2+\frac12\kappa_2\dot{q}_2^2 -\frac{C}{4\pi}\epsilon^{ab} q_a\dot{q}_b
\end{equation}

where $q_{1(2)}=\int d^2x\,\varphi_{1(2)}(x,t)$ are the zero-momentum components of the Goldstone fields, and we have eliminated cross-terms like $\dot{q}_1\dot{q}_2$ by a suitable field redefinition. The canonical momenta $\frac{\partial L}{\partial\dot{q}_a}$ are

$$p_1=\kappa_1\dot{q}_1+\frac{C}{4\pi}q_2,\,\,\,\, p_2=\kappa_2\dot{q}_2-\frac{C}{4\pi} q_1$$

with the usual commutation relation $[q_a,p_b]=i\delta_{ab}$ imposed in canonical quantization. The expression for the conserved dipole charges under the shift symmetries $q_a\to q_a+c_a$ are:

$${Q}_1=\kappa_1\dot{q}_1+\frac{C}{2\pi}q_2=p_1-\frac{C}{4\pi}q_2,\,\,\,\,{Q}_2=\kappa_2\dot{q}_2-\frac{C}{2\pi}q_1=p_2+\frac{C}{4\pi}q_1$$

When converted to operators, using the canonical commutation relations lets us easily compute $[Q_a,Q_b] = -i\frac{C}{2\pi}\epsilon^{ab}$.

\section{Quantization of the Projected Dipole Charge Commutator}
\label{App:Quantization}

Here, we will prove that the commutator of projected dipole charge generators $[\hat{Q}^a,\hat{Q}^b]=-i\frac{C}{2\pi}$ is quantized with $C$ an integer. Morally speaking, this result holds for the same reason that the many-body Hall conductance is quantized in 2-D so long as the system is uniquely gapped\footnote{Here, we refer to a system as "uniquely gapped" if the gapped ground state is nondegenerate in the thermodynamic limit.}, and our proof will follow almost the exact same steps as in the work of Niu-Thouless-Wu \cite{Niu_Kosterlitz_Wu_ManyBodyChern,wen_qft_bible}. We will assume that the system lives on a periodic square lattice with $N_x\times N_y$ sites, and that the theory can be formulated as dipolar Goldstone modes coupled to a 'matter' sector of the theory which is uniquely gapped (but may be interacting). Note that the commutator of our dipole charges is dimensionless because we are working with a rescaled dipole charge operator $\hat{Q}_a = \frac{1}{a}\sum_i(\hat{a}\cdot\vec{r}_i)\hat{n}_i$; in the field theory context, this simply amounts to a rescaling of the $\varphi_a$-field. \hfill \break

Because the matter sector is gapped, we may integrate it out to recover the effective Goldstone theory, whose zero-momentum fluctuations will be described by the quadratic Lagrangian
$$L_{goldstone} =\frac{1}{N_xN_y}\biggr(-\frac{C}{4\pi}\epsilon^{ab}\varphi_a\partial_t\varphi_b + \kappa^{ab}\partial_t\varphi_a\partial_t\varphi_b\biggr)$$
where we regularize the system on a finite number $N_x\times N_y$ of sites. As explained in App. \ref{App:CommutatorFromAction}, this Goldstone effective action yields the relation $[\hat{Q}^a,\hat{Q}^b]=-i\frac{C}{2\pi}$, so we now need to prove that the coefficient $C$ in the Lagrangian is quantized. 

We will consider particular time-dependent fluctuations of the fields $\varphi_a = \theta_a(t)/N_a$, where the frequency scale of the time variation is much larger than the many-body gap in the matter sector; i.e. the adiabatic theorem applies to the time evolution of the matter sector when the variation of the Goldstone fields is treated as an external drive. Such a variation between times $[t,t+\Delta t]$ may be affected by acting on the mean-field ground state with a symmetry operator
$$\ket{GS} \to e^{-i\hat{Q}_a\partial_t\theta_a\Delta t/N_a}\ket{GS} = \ket{GS;\theta_a(t)}$$
because this is equivalent to transforming the fermionic lattice operators as
$$\hat{\psi}_i \to e^{i\hat{Q}_a\partial_t\theta_a\Delta t/N_a}\hat{\psi}_ie^{-i\hat{Q}_a\partial_t\theta_a\Delta t/N_a} = \hat{\psi}_ie^{ir^a_i\partial_t\theta_a\Delta t/L_a}$$
where $L_a=aN_a$ is the system size; in other words, the Goldstone field is parametrized as a symmetry rotation of the fermionic operator. Integrating this transformation, we have fermionic operators following the variation of the Goldstone field as $\hat{\psi}_i \to  \hat{\psi}_ie^{ir^a_i\theta_a(t)/L_a}$. The periodicity in space identifies sites $N_a+1\cong 1$, and after adding the Goldstone fluctuation we have $\hat{\psi}_{1,i}  = e^{i\theta_x(t)}\psi_{L_a+1,i}$, i.e. the originally periodic boundary condition $\hat{\psi}_{1,i}  = \psi_{L+1,i}$ has become twisted (the same is true for the $y$ direction). Because the Goldstone fluctuation is spatially uniform and the evolution of the Goldstone field from $0\to\theta_a(t)$ is adiabatic, this state with the twisted boundary condition is the ground state $\ket{GS;\theta_a(t)}$ of a new mean-field Hamiltonian with parametric dependence on the two Goldstone fields $\theta_x,\theta_y$. We also see that the twisted boundary conditions described by $\theta_a$ and $\theta_a+2\pi$ are the same. Accordingly, the states $\ket{GS;\theta_a(t)}$ are ground states of a Hamiltonian $\hat{H}(\varphi^a,\varphi^b)$ parametrized by coordinates $\theta_a(t)=N_a\varphi_a(t)$ on the 2-torus $[0,2\pi]^2$. 

It is well-known that the Berry phase due to evolving the ground state $\ket{GS;\theta_a(t)}$ along the path $\gamma(t)$, given by $(\theta_x(t),\theta_y(t))=(0,0)\to(2\pi,0)\to(2\pi,2\pi)\to(0,2\pi)\to (0,0)$, is quantized to be in $2\pi\mathbb{Z}$; this is because such a path is equivalent to a stationary point on the 2-torus, and thus its Berry phase must vanish modulo $2\pi$. 

Now, we relate the Berry phase along this path to the term proportional to $C$ in the Goldstone effective action. Following \cite{berry_phase_effective_action}, we identify the effective action with the phase accumulated by the ground states $\ket{GS,\theta_a}$ 
$$S_{eff} = \int dtL_{goldstone}[\varphi_a(t)] = \int dt\bra{GS,\theta_a(t)}i\partial_t-\hat{H}(\varphi^a,\varphi^b)\ket{GS,\theta_a(t)} \sim \int dt\bra{GS,\theta_a(t)}i\partial_t\ket{GS,\theta_a(t)} + TE_{G.S.}$$
where $T$ is the time interval of the path $\gamma(t)$ and $E_{G.S.}$ is the ground state energy of the untwisted system (which may be set to zero), and the thermodynamic limit was taken to ensure that $E_{G.S.}$ does not depend on the twisting. In other words, the effective Goldstone action for spatially uniform, slowly-varing Goldstone fields is exactly equal to the Berry phase of the mean-field ground state, which was discussed above to be quantized along the path $\gamma(t)$. Finally, we may evaluate the effective action along the path $\gamma(t)$:

$$S_{eff}[\gamma] = \int dt\,\biggr(-\frac{C}{4\pi}\epsilon^{ab}\theta_a(t)\partial_t\theta_b(t) + \kappa^{ab}\partial_t\theta_a\partial_t\theta_b\biggr) $$

The second term vanishes as the variation of the fields is taken to be adiabatic, and the first term is simply $-\frac{C}{2\pi}$ times the area $(2\pi)^2$ of the parameter space in $(\theta_x,\theta_y)$ swept out by the path $\gamma$, by Green's theorem. Accordingly, $S_{eff}[\gamma]  = -2\pi C$, and per the prior discussion this establishes $C$ as an integer. Of course, this integer $C$ is the Chern number of the matter sector, since that is precisely the quantized Berry phase associated to the space of twisted boundary conditions, per \cite{Niu_Kosterlitz_Wu_ManyBodyChern}.

One could understand this construction more abstractly by appealing to the discussion in \cite{Watanabe_Goldstone_Theorem}, where it is shown that if $\varphi^a,\varphi^b$ are Goldstone fields for a $U(1)\times U(1)$ symmetry, the quantization of the coefficient $C$ stems from the compactness of the group manifold. We however caution that the dipolar symmetry operator does not merely belong to an internal $U(1)\times U(1)$ subgroup of our lattice system but is part of a more general multipole algebra that includes nontrivial commutation relations with translation \cite{Gromov_Multipole}. Furthermore, the previous proof shows that the connection between the quantized term in the Goldstone action and the mean-field topology is particular to the case of dipolar Goldstones, and the role of the mean-field topology is crucial to the rest of our work as it dictates the nature of phase transitions between Goldstone field theories with differing $C$.

\section{Simplification of the RPA Goldstone Propagator}
\label{App:RPAGoldstone}

In this section, we justify the low-energy form of the RPA Goldstone propagator as given in the main text. The first order of business is to compute the polarization tensor for the free Dirac fermions, since the Goldstone self-energy tensor at one-loop is nothing but the spacelike part of this polarization tensor. This will have the structure
\begin{equation}\label{eq:General_Dirac_Polarization}
\Pi^{ab} = \Pi_e(p,\omega)\frac{1}{(\omega^2+p^2)^{1/2}}(\delta_{ab}(\omega^2+p^2) - p_ap_b) + \Pi_o(p,\omega)\epsilon_{ab}\omega
\end{equation}

On symmetry considerations, we can see that $\Pi_o=\sign(M)\frac{g^2}{2}\frac{1}{4\pi}$, with $\Pi_o$ vanishing at one loop when $M=0$. At one-loop order, and at $M=0$, $\Pi_e(p,\omega)=-\frac{1}{16}$ \cite{mpa_fisher_mott_transition_anyon_gas}. A convenient basis to express the Goldstone components in is the transverse/longitudinal basis, where the transverse component $\varphi_T^a(p)$ is defined by $p_a\varphi_T^a(p)=0$ and the longitudinal component $\varphi_L^a(p)$ is polarized parallel to its momentum $p$. The one-loop inverse propagator for the Goldstones in this transverse/longitudinal basis is given by:

\begin{equation}\label{eq:exact_RPA_goldstone_inverse_propagator}
D_{ab}^{-1} = \begin{pmatrix}(\omega^2+v_T^2\mathbf{p}^2)+\frac{g^2}{8}\sqrt{\omega^2+\mathbf{p}^2} & \frac12\frac{g^2}{4\pi}\omega \\ -\frac12\frac{g^2}{4\pi}\omega & (\omega^2+v_L^2\mathbf{p}^2)+\frac{g^2}{8}\frac{\omega^2}{\sqrt{\omega^2+\mathbf{p}^2}}\end{pmatrix}
\end{equation}

We see that only the half-quantized Berry phase term for the Goldstones couples the transverse and longitudinal modes; if we relax the assumption of rotational symmetry, anisotropy will also couple these two modes. To extract the most relevant information from this propagator, we inspect the singularities of the propagator $D_{ab}$. For small momentum $p$, we find a singularity $\omega\sim p^{3/2}$, indicating a dynamical exponent of $z=3/2$. Using this new scaling to systematically discard all but the most RG-relevant terms, the form of the Goldstone propagator near this pole is 

\begin{equation}\label{eq:Longitudinal_32_pole_APP}
D^L_{ab}(\omega,\mathbf{p}) \sim \lambda^2\frac{p}{ \omega^2+\zeta^2v_L^2p^3 }\times \begin{pmatrix}0 & 0 \\ 0 &     1 \end{pmatrix}  
\end{equation}
which is a purely longitudinal pole; here $\lambda^2=\frac{8g^{-2}}{1+1/\pi^2}$ and $\zeta^2 = \frac{8}{g^2(1+1/\pi^2)}v_L^2$. There is another singularity at $\omega\sim p$; the most singular and relevant part of the propagator in the vicinity of this singularity is given by
\begin{equation}\label{eq:Transverse_linear_pole_APP}
D^T_{ab}(\omega, \mathbf{p}) \sim \lambda^2\frac{1}{\sqrt{\omega^2+p^2}} \begin{pmatrix}1   & 0\\ 0 &  0 \end{pmatrix} ,
\end{equation}
which is purely transverse, and notably implies a higher scaling dimension for the transverse component of the Goldstones as compared to the longitudinal component.

We can understand the behavior of the longitudinal Goldstone boson by studying the Lagrangian more carefully. The Goldstone-Fermion coupling takes the form $\varphi_a J^a$ where $J^a$ is the spacelike part of the conserved $U(1)$ current, which obeys the conservation law $\omega J^0 =- \vec{p}\cdot\vec{J}$. Since the longitudinal Goldstone boson $\varphi_L = p_a\varphi^a/\abs{p}$, the interaction term between the longitudinal Goldstone and Fermions can be taken to be $\varphi_L \frac{\omega}{p}\bar{\psi}\gamma^0\psi$, where $\bar{\psi}\gamma^0\psi$ is the $U(1)$ charge density. Hence the exact self-energy for the Longitudinal Goldstone will be

\begin{equation}\label{eq:Formal_Long_Boson_Self_Energy}
\Pi_L(\omega,\mathbf{p}) = \frac{\omega^2}{p^2}\avg{J^0(\omega,\mathbf{p})J^0(-\omega,-\mathbf{p})}_{c}
\end{equation}
where the subscript-$c$ indicates a connected correlation function. At one loop, a standard evaluation of the charge correlator yields

$$\propto \frac{\omega^2}{p^2}\frac{p^2}{\sqrt{\omega^2+p^2}}\,\underset{z=3/2}{\longrightarrow}\,\frac{\omega^2}{p}$$
as we have just shown in the computation of $D^L_{ab}(\omega,\mathbf{p})$.

\section{One-Loop Fermion Self-Energy}\label{App:OneLoopFermionSE}

Now, we derive the Fermion self-energy $\Sigma(\nu,\mathbf{q})$ at one-loop using the RPA-corrected Goldstone propagator, given by Eq. \ref{eq:Bare_Fermion_Self_Energy} in the main text. The most relevant contribution is from the $\omega\sim p^{3/2}$ in the corrected Goldstone propagator:

$$\Sigma(\nu,\mathbf{q}) = -\frac{\lambda^2g^2}{N}\int \frac{d\omega d^2p}{(2\pi)^3} \frac{1}{ \omega^2+\zeta^2\mathbf{p}^3 }  ((\omega+\nu)\fsh{\mathbf{p}}\gamma^0\fsh{\mathbf{p}})+(\fsh{\mathbf{p}}(\fsh{\mathbf{p}}+\fsh{\mathbf{q}})\fsh{\mathbf{p}}))\frac{p^{-1}}{(\omega+\nu)^2 + (\mathbf{p}+\mathbf{q})^2}$$

We can evaluate this to leading order in $\nu,\mathbf{q}$ by considering nonzero $\nu$ and nonzero $\mathbf{q}$ separately. First we handle the case of $\mathbf{q}=0$:

$$\Sigma(\nu,0) = \gamma^0\frac{\lambda^2g^2}{N}\frac{1}{(2\pi)^3}\int d\omega d^2p \frac{\omega}{ (\omega-\nu)^2+\zeta^2p^3 }  \frac{p}{\omega^2 + p^2}$$

Evaluating the contour integral over $\omega$,

$$ = \gamma^0\cdot 2\pi i\frac{\lambda^2g^2}{N}\frac{1}{(2\pi)^3}\int d^2p \biggr(  \frac{1}{2}\frac{p}{ (ip-\nu)^2+\zeta^2p^3 }  + \frac{(\nu +i\zeta p^{3/2})}{2i\zeta p^{1/2} }  \frac{1}{(\nu +i\zeta p^{3/2})^2 + p^2}\biggr)$$

The first integral can be scaled with $p\to p/\nu$ and the second integral naturally scales as $p\to p/\nu^{2/3}$. Accordingly,
$$ = \gamma^0\cdot 2\pi i\frac{\lambda^2g^2}{N}\frac{1}{(2\pi)^3}\biggr(\int d^2x \frac{1}{2}\nu\frac{x}{ (ix-1)^2+\zeta^2\nu x^3 }  + \int d^2x\frac{(1 +i\zeta x^{3/2})}{2i\zeta x^{1/2} }  \frac{1}{(1 +i\zeta x^{3/2})^2 + \nu^{-2/3}x^2}\biggr)$$

When a UV cutoff is implemented, we see that the first integral only contributes an $O(\nu)$ and $O(1)$ piece to the self-energy. (Clearly, the $O(1)$ piece will always cancel out, so we only turn our attention to contributions that vanish as $\nu\to 0$.) The second integral must be further rescaled with $y = x\nu^{-1/3}$:

$$ \gamma^0\frac{\pi \lambda^2g^2}{\zeta N}\frac{1}{(2\pi)^2}\nu^{1/2}\int dy\,y^{1/2} (1 +i\zeta \nu^{1/2}y^{3/2})  \frac{1}{(1 +i\zeta \nu^{1/2}y^{3/2})^2 + y^2}$$

In the small-$\nu$ limit, this contains a $O(\nu^{1/2})$ contribution and an $O(1)$ contribution which cancels the $O(1)$ part of the previous integral. Thus,

$$\Sigma^{(1)}(0,\nu) = \gamma^0\frac{\pi \lambda^2 g^2}{\zeta N}\frac{1}{(2\pi)^2}\nu^{1/2}\int dy\,\frac{y^{1/2}}{1+y^2} + \ldots = \gamma^0\frac{\lambda^2 g^2}{4\sqrt{2}\zeta N}\nu^{1/2} + \ldots$$

Now we focus on the case $\Sigma^{(1)}(\mathbf{q},0)$. It is convenient to compute 

$$\frac12\Tr(\fsh{\mathbf{q}}\Sigma^{(1)}(\mathbf{q},0)) = -\frac{\lambda^2 g^2}{2N}\frac{1}{(2\pi)^3}\int d\omega d^2p \frac{1}{ \omega^2+\zeta^2\mathbf{p}^3 }  \Tr(\fsh{\mathbf{q}}\fsh{\mathbf{p}}(\fsh{\mathbf{p}}+\fsh{\mathbf{q}})\fsh{\mathbf{p}})\frac{p^{-1}}{\omega^2 + (\mathbf{p}+\mathbf{q})^2}$$

$$= -\frac{\lambda^2 g^2}{2\zeta N }\frac{\pi}{(2\pi)^3}\int  \frac{d^2p}{p^{5/2}}  \frac{1}{\zeta p^{3/2}\abs{\mathbf{p}+\mathbf{q}} + p^2+q^2+2(\vec{p}\cdot\vec{q}) }\biggr(4(\vec{q}\cdot\vec{p})^2-2q^2p^2+2p^2(\vec{q}\cdot\vec{p})\biggr)$$

We now take the perspective that a singular self-energy for the fermions can only be generated by interaction with the soft boson modes; thus, we evaluate the above integral only for momenta $p\ll q$, and expand

$$\abs{\mathbf{p}+\mathbf{q}} =q + \frac{\vec{p}\cdot\vec{q}}{q} + \frac{p^2}{2q} - \frac{(\vec{p}\cdot\vec{q})^2}{2q^3} + O(p^3)$$

Nondimensionalizing $p\to p/q=x$, and taking $\vec{p}\cdot\vec{q} = q^2x\cos\theta$,
$$\implies \frac{1}{q}\frac12\Tr(\fsh{\mathbf{q}}\Sigma^{(1)}(\mathbf{q},0))\sim -\frac{\lambda^2 g^2}{2\zeta N }\frac{\pi}{(2\pi)^3}q^{1/2}\int  \frac{d^2x}{x^{5/2}}  \frac{\biggr(4x^2\cos^2\theta-2x^2+2x^3\cos\theta\biggr)}{\zeta q^{1/2}x^{3/2}\biggr(1 + x\cos\theta + \frac12 x^2 - \frac12 x^2\cos^2\theta\biggr) + x^2+1+2x\cos\theta }$$

We wish to evaluate the above integral for small $q$; thus we may expand

$$\implies \frac{1}{q}\frac12\Tr(\fsh{\mathbf{q}}\Sigma^{(1)}(\mathbf{q},0))\sim -\frac{\lambda^2 g^2}{2\zeta N }\frac{\pi}{(2\pi)^3}\int  \frac{d^2x}{x^{5/2}}  \frac{4x^2\cos^2\theta-2x^2+2x^3\cos\theta}{1+2x\cos\theta+x^2}\biggr(q^{1/2} - q\frac{\zeta x^{3/2}}{1+x^2+2x\cos\theta }\biggr) $$

Na\"ively, we would conclude that $\Sigma(\mathbf{q},0)\sim \abs{q}^{1/2}\gamma_{\vec{q}}$. But in fact, the first term in the above parentheses integrates to zero. Accordingly, the low-$p$ contribution to the self-energy results in a term 

$$\Sigma(\mathbf{q},0) = c\fsh{\mathbf{q}} + O\ldots$$

where $c$ is a constant that depends on $\lambda,\zeta$ and other numerical factors. We do not concern ourselves with the form of this constant $c$, since the subsequent discussion in the main text and the next Appendix \ref{App:FermionSCSE} shows that this contribution $c\fsh{\mathbf{q}}$ to the self-energy is irrelevant compared to higher-loop corrections.

\section{Self-Consistent Self-Energies}
\label{App:FermionSCSE}

To begin, we derive the form of the fermion self-energy presented in Sec. \ref{SubSec:SC_Fermion_Self_Energy}. We begin with the ansatz 
\begin{equation}
    \label{eq:SEAnsatz_App}
    \Sigma(\nu,\vec{q}) = Z_0\abs{\nu}^{1/2}\sign(\nu)\gamma^0+Z_1\zeta^{1/2}\abs{q}^{3/4}\gamma_{\vec{q}} + \ldots
\end{equation}
where the purported non-analyticity is motivated by the results of a one-loop calculation (App. \ref{App:OneLoopFermionSE}), and impose the self-consistent condition (see also Fig. \ref{fig:Self_consistent_fermion})

\begin{equation}
    \label{eq:Self_Consistency_App}
\Sigma(\nu,\vec{q}) = -\frac{\lambda^2g^2}{N}\int \frac{d\omega d^2p}{(2\pi)^3} \frac{1}{ \omega^2+\zeta^2\mathbf{p}^3 }\frac{p_ap_b}{p}  \frac{\gamma^a(Z_0\abs{\omega+\nu}^{1/2}\sign(\omega+\nu)\gamma^0+\zeta^{1/2} Z_1\abs{p+q}^{3/4}\gamma_{\vec{p}+\vec{q}})\gamma^b}{Z_0^2\abs{\omega+\nu}+\zeta Z_1^2\abs{p+q}^{3/2}}.
\end{equation}


In order to recover the ansatz \ref{eq:SEAnsatz_App}, we can evaluate the integral \ref{eq:Self_Consistency_App} with either $\vec{q}=0$ or $\omega=0$; comparison with \ref{eq:SEAnsatz_App} will yield two self-consistency equations that fix $Z_0,Z_1$. First setting $\vec{q}=0$, we have
\begin{equation}
\label{eq:Self-Consistency_nu_1}
    \frac12\Tr(\gamma^0\Sigma(\nu,0)) = \frac{\lambda^2g^2}{N}\int \frac{d\omega d^2p}{(2\pi)^3}  \frac{p}{ (\omega-\nu)^2+\zeta^2\mathbf{p}^3 }  \frac{Z_0\abs{\omega}^{1/2}\sign(\omega)}{Z_0^2\abs{\omega}+Z_1^2 \zeta \abs{p}^{3/2}}
\end{equation}

Passing to spherical coordinates and defining $x=\zeta^{2/3} p/\abs{\nu}^{2/3}$, $s=\omega/\abs{\nu}$,

\begin{equation}
\label{eq:Self-Consistency_nu_2}
    \frac12\Tr(\gamma^0\Sigma(\nu,0)) = \frac{{\lambda^2}g^2\zeta^{-2}}{(2\pi)^2N}\abs{\nu}^{1/2}\int_{-\infty}^\infty ds\int_0^\infty dx\,  \frac{x^2}{ (s-1)^2+x^3 }  \frac{Z_0\abs{s}^{1/2}\sign(s)}{Z_0^2\abs{s}+ Z_1^2\abs{x}^{3/2}}
\end{equation}

Now we compute

\begin{equation}
\label{eq:Self-Consistency_q}
\frac12\Tr(\gamma_{\vec{q}}\Sigma(0,\vec{q}))  = -\frac{\lambda^2g^2}{N}\int \frac{d\omega d^2p}{(2\pi)^3}  \frac{1}{ \omega^2+\zeta^2\mathbf{p}^3 }\frac{p_ap_b}{\abs{\mathbf{p}}} \zeta^{1/2}Z_1\abs{\mathbf{p}+\mathbf{q}}^{3/4} \frac{\Tr(\gamma_{\vec{q}}\gamma^a\gamma_{\vec{p}+\vec{q}}\gamma^b)}{Z_0^2\abs{\omega}+Z_1^2\zeta\abs{\mathbf{p}+\mathbf{q}}^{3/2}}
\end{equation}

Defining $s = \omega/(\zeta q^{3/2})$, $\vec{r}=\vec{p}/q$, and w.l.o.g. taking $\vec{q}=q\hat{x}$,

\begin{equation}
\label{eq:Self-Consistency_q_1}
\frac12\Tr(\gamma_{\vec{q}}\Sigma(0,\vec{q}))  = -\frac{\lambda^2g^2\zeta^{-3/2}}{N}\int \frac{ds d^2x}{(2\pi)^3}  \frac{1}{ s^2+\mathbf{r}^3 }\frac{r_ar_b}{\abs{\mathbf{r}}} Z_1\abs{\mathbf{r}+\hat{x}}^{3/4} \frac{\Tr(\gamma_{\hat{x}}\gamma^a\gamma_{\vec{r}+\hat{x}}\gamma^b)}{Z_0^2\abs{s}+Z_1^2\abs{\mathbf{r}+\hat{x}}^{3/2}}
\end{equation}

Upon using the ansatz Eq. \ref{eq:SEAnsatz_App} to compute $\frac12\Tr(\gamma^{0,i}\sum(\omega,0))$, Eqs. \ref{eq:Self-Consistency_nu_2} and \ref{eq:Self-Consistency_q_1} turn into two self-consistency equations:

\begin{equation}
\label{eq:Self-Consistency_Dimless}
\begin{split}
    1 &= \frac{{\lambda^2}g^2\zeta^{-2}}{(2\pi)^2N}\int_{-\infty}^\infty ds\int_0^\infty dx\,  \frac{x^2}{ (s-1)^2+x^3 }  \frac{\abs{s}^{1/2}\sign(s)}{Z_0^2\abs{s}+ Z_1^2\abs{x}^{3/2}}\\
    1 &= -\frac{\lambda^2g^2\zeta^{-2}}{(2\pi)^3N}\int ds d^2x\,  \frac{\Tr(\gamma_{\hat{x}}\gamma^a\gamma_{\vec{r}+\hat{x}}\gamma^b)}{ s^2+\mathbf{r}^3 }\frac{r_ar_b}{\abs{\mathbf{r}}} \frac{\abs{\mathbf{r}+\hat{x}}^{3/4} }{Z_0^2\abs{s}+Z_1^2\abs{\mathbf{r}+\hat{x}}^{3/2}}
\end{split}
\end{equation}

If we write $Z_0 = c_1\lambda g/\zeta \sqrt{N}$ and $Z_1 = c_2\lambda g/\zeta\sqrt{N}$, these equations become 

\begin{equation}
\label{eq:Self-Consistency_Dimless}
\begin{split}
    1 &= \frac{1}{(2\pi)^2}\int_{-\infty}^\infty ds\int_0^\infty dx\,  \frac{x^2}{ (s-1)^2+x^3 }  \frac{\abs{s}^{1/2}\sign(s)}{c_1^2\abs{s}+ c_2^2\abs{x}^{3/2}}\\
    1 &= -\frac{1}{(2\pi)^3}\int ds d^2r\,  \frac{\Tr(\gamma_{\hat{x}}\gamma^a\gamma_{\vec{r}+\hat{x}}\gamma^b)}{ s^2+\mathbf{r}^3 }\frac{r_ar_b}{\abs{\mathbf{r}}} \frac{\abs{\mathbf{r}+\hat{x}}^{3/4} }{c_1^2\abs{s}+c_2^2\abs{\mathbf{r}+\hat{x}}^{3/2}}
\end{split}
\end{equation}

The constants $c_1$ and $c_2$ are determined by the transcendental equations given by Eq. \ref{eq:Self-Consistency_Dimless} and are universal constants; i.e. they do not depend on the parameters $\lambda$ or $\zeta$.

Now that we have self-consistently fixed the fermion self-energy by resumming rainbow diagrams, we can compute corrections to the boson self-energy beyond one-loop by summing polarization bubbles using the corrected fermion propagator:

$$\Pi(\omega,\vec{q}) = N\frac{\zeta^2}{\lambda^2} \int d\nu d^2p\frac{2c_1^2\abs{\nu}^{1/2}\abs{\nu-\omega}^{1/2} - \frac{p_ap_b}{p^2}c_2^2\zeta\abs{p}^{3/4}\abs{p+q}^{3/4}\Tr(\gamma^a \gamma_{\vec{p}}\gamma^b\gamma_{\vec{p}+\vec{q}})}{(c_1^2\abs{\nu} + c_2^2\zeta p^{3/2})(c_1^2\abs{\nu-\omega} + c_2^2\zeta \abs{p-q}^{3/2})}$$
Nondimensionalizing the above integral by writing $s = \nu/\abs{\omega}$, $\vec{r}=\vec{p}/\abs{q}$, we see that the self-energy takes the form



$$ \Pi(\omega,\vec{q}) = N\frac{\zeta^2}{\lambda^2}q^2 F\biggr(\frac{\abs{\omega}}{\zeta q^{3/2}}\biggr)$$
where the function $F(\cdot)$ is dimensionless and independent of $N$, and depends on the constants $c_1,c_2$.

\end{document}